\documentclass{bmcart}

\usepackage[utf8]{inputenc} 

\usepackage{hyperref}
\usepackage{amssymb}
\usepackage{hyperref}
\usepackage{booktabs}
\usepackage{amsmath}
\usepackage{placeins}
\usepackage{changepage}
\usepackage{graphicx}

\usepackage{soul} 

\startlocaldefs
\endlocaldefs

\begin{document}

\begin{frontmatter}

\begin{fmbox}
\dochead{Research}


\title{Diversity dilemmas: uncovering gender and nationality biases in graduate admissions across top North American computer science programs}


\author[
   addressref={aff1}, 
   corref={aff1}, 
   email={kalhor.ghazal@ut.ac.ir}  
]{\inits{GK}\fnm{Ghazal} \snm{Kalhor}}

\author[
   addressref={aff1},
   email={tzeraati@ut.ac.ir}
]{\inits{TZ}\fnm{Tanin} \snm{Zeraati}}

\author[
   addressref={aff2},   
   email={b.bahrak@teias.institute} 
]{\inits{BB}\fnm{Behnam} \snm{Bahrak}}


\address[id=aff1]{
  \orgname{School of Electrical and Computer Engineering, College of Engineering, University of Tehran}, 
  \city{Tehran},                              
  \cny{Iran}                                    
}

\address[id=aff2]{
  \orgname{Tehran Institute for Advanced Studies}, 
  \city{Tehran},                              
  \cny{Iran}                                    
}



\end{fmbox}


\begin{abstractbox}

\begin{abstract} 
Although different organizations have defined policies towards diversity in academia, many argue that minorities are still disadvantaged in university admissions due to biases. Extensive research has been conducted on detecting partiality patterns in the academic community. However, in the last few decades, limited research has focused on assessing gender and nationality biases in graduate admission results of universities. In this study, we collected a novel and comprehensive dataset containing information on approximately 14,000 graduate students majoring in computer science (CS) at the top 25 North American universities. We used statistical hypothesis tests to determine whether there is a preference for students’ gender and nationality in the admission processes. In addition to partiality patterns, we discuss the relationship between gender/nationality diversity and the scientific achievements of research teams. Consistent with previous studies, our findings show that there is no gender bias in the admission of graduate students to research groups, but we observed bias based on students’ nationality. 

\end{abstract}


\begin{keyword}
\kwd{graduate admission}
\kwd{statistical analysis}
\kwd{partiality patterns}
\kwd{gender equality}
\kwd{nationality diversity}
\end{keyword}


\end{abstractbox}
%

\end{frontmatter}



\section*{Introduction}
Every year, many students from all over the world apply to pursue their graduate studies at top universities in North America~\cite{sharaievska2019we}. Despite the committee-based nature of admission to many of these universities, professors still play a prominent role in accepting students and providing them with financial support~\cite{posselt2014toward}. As a result, students often directly contact faculty members to enhance their chances of admission. Furthermore, students who are admitted by a committee must find an academic advisor and research group, and faculty members have the authority to approve or reject these requests. Consequently, their research group may demonstrate a preference for accepting students of similar gender, country of origin, or previous universities. In this study, we aim to examine the existing biases in interactions with computer science faculty members at top North American universities and their preferences regarding nationality and gender when selecting graduate students for their research group.

In addition to establishing fair admission systems, it is crucial to enhance diversity in academia. Promoting diversity within universities enables them to have a greater impact on societies~\cite{bollinger2007diversity}. This is because institutions aim to address social issues, which cannot be effectively achieved without embracing diversity~\cite{Smith2020DiversitysPF}. Furthermore, it is argued that being in a diverse environment can broaden students' horizons~\cite{maruyama2000university}.

Most prestigious universities typically strive to ensure fairness in the admission process for their graduate programs. Various factors, such as merit, gender equality, and diversity, contribute to establishing a fair graduate admission system~\cite{pitman2016understanding, posselt2014toward}. However, it is argued that admitting a greater number of marginalized students for graduate education at U.S. universities remains a contentious issue~\cite{Barrera2006MakingUG}.

To the best of our knowledge, only three studies have focused on assessing gender or nationality bias in graduate admissions, and all of them were conducted prior to 2000. Bickel and Hammel~\cite{bickel1975sex} analyzed admission results from various schools at the University of California, Berkeley to examine the presence of a gender gap. They found statistically significant favoritism towards female applicants. Maxwell and Jones~\cite{Maxwell1976} employed adjustment techniques to compare admission rates between women and men in four graduate programs at the University of North Carolina, Chapel Hill. Their findings suggested that gender was not a significant factor in admission decisions. Subsequently, the authors of~\cite{attiyeh1997testing} discussed the influence of demographic attributes, such as gender and country of citizenship, on graduate admission decisions at top-ranked American universities. Their results indicated that these universities placed greater emphasis on admitting U.S. students, and female applicants received some degree of preference. Our work builds upon these studies by addressing questions regarding gender/nationality bias in more recent and comprehensive graduate admissions data. The dataset we collected for this study encompasses a larger number of students and includes a greater number of universities. 

Some studies have examined the impact of gender/nationality diversity on the performance of research teams. In~\cite{barjak2008international}, the authors investigated the level of cultural diversity at which a research group achieves the highest performance. AlShebli et al.~\cite{AlShebli2018} analyzed author lists of research papers to explore the influence of diversity in characteristics such as gender and ethnicity on the success of research teams. Llorens et al.~\cite{llorens2021gender} demonstrated the existence of gender bias throughout scholars' academic careers, affecting aspects such as career opportunities, promotion, and grant allocation. They also proposed solutions at various levels to enhance diversity, highlighting its importance for scientific success. The authors of~\cite{nielsen2018making} examined different facets of gender diversity and reported its positive impact on creativity and performance in scientific domains. Kamerlin~\cite{Kamerlin2020} addressed bias issues in academia and presented strategies to promote gender diversity in academic environments. Powell~\cite{powell2018these} utilized citation count to quantify the success of research papers and investigated its relationship with various aspects of diversity, such as gender, age, ethnicity, and affiliation, among the authors. In addition to citation count, we consider faculty members' h-index and publication count as measures of success for their research groups.

Many initiatives have been undertaken to enhance diversity in computer science. The author of~\cite{Larsen2005} emphasized that these efforts should not be limited to achieving gender equity alone. Wilson~\cite{Wilson2014} highlighted how his team in Hour of Code decided to translate their lectures into multiple languages and establish branches in more countries to promote diversity in computer science. One of the primary objectives of their program is to globalize computer science~\cite{Partovi2015}. Increasing students' awareness of diversity and inclusion is a crucial step towards fostering a more diverse community of computer scientists~\cite{GarciaHolgado2019}. These studies collectively underscore the significance of addressing diversity issues in academia.

In this study, we aim to address the following questions:
\begin{itemize}
\item Do professors exhibit a preference for admitting students of the same gender to their research group?
\item Are they inclined to accept students who share their country of origin?
\item How do these bias patterns evolve over time?
\item Is there any correlation between the diversity of gender or nationality among team members and the research team's productivity?
\end{itemize}

Our contributions can be summarized as follows:
\begin{enumerate}
\item We provide a comprehensive description of the dataset collected for this study, highlighting its various features.
\item We analyze the gender distributions of students and faculty members and conduct hypothesis tests to examine the presence of gender bias in the selection of students for graduate study.
\item We investigate the distributions of advisors and students' home countries and explore the existence of bias in this variable.
\item We construct an advisor-student relationship network using our dataset and calculate centrality metrics to identify the most influential countries in higher education.
\item We examine the trends in gender/nationality biases and diversities among advisor-student pairs over time using Mann-Kendall tests.
\item We assess the correlations between academic success and diversity measures to analyze the relationship between gender/nationality diversity and the performance of research groups.
\end{enumerate}

The rest of this paper is structured as follows. The "Materials and methods" section offers an overview of the data collection process and delineates the diverse features within our dataset. Following this, our discoveries are outlined and analyzed in the "Results and discussion" section. In the "Future work" section, potential directions for future research are proposed. Finally, the "Conclusion" section succinctly summarizes the key takeaways of the paper.

\section*{Materials and methods}
In this section, we define the techniques and metrics that we use in answering our research questions. Moreover, we describe the dataset that we collected for this study.

\subsection*{Methods}
In this part, we introduce the algorithms and statistical tests utilized in our study.

\subsubsection*{Disparity filter}
\begin{figure}[!ht]
\centering
\includegraphics[width=\hsize, keepaspectratio]{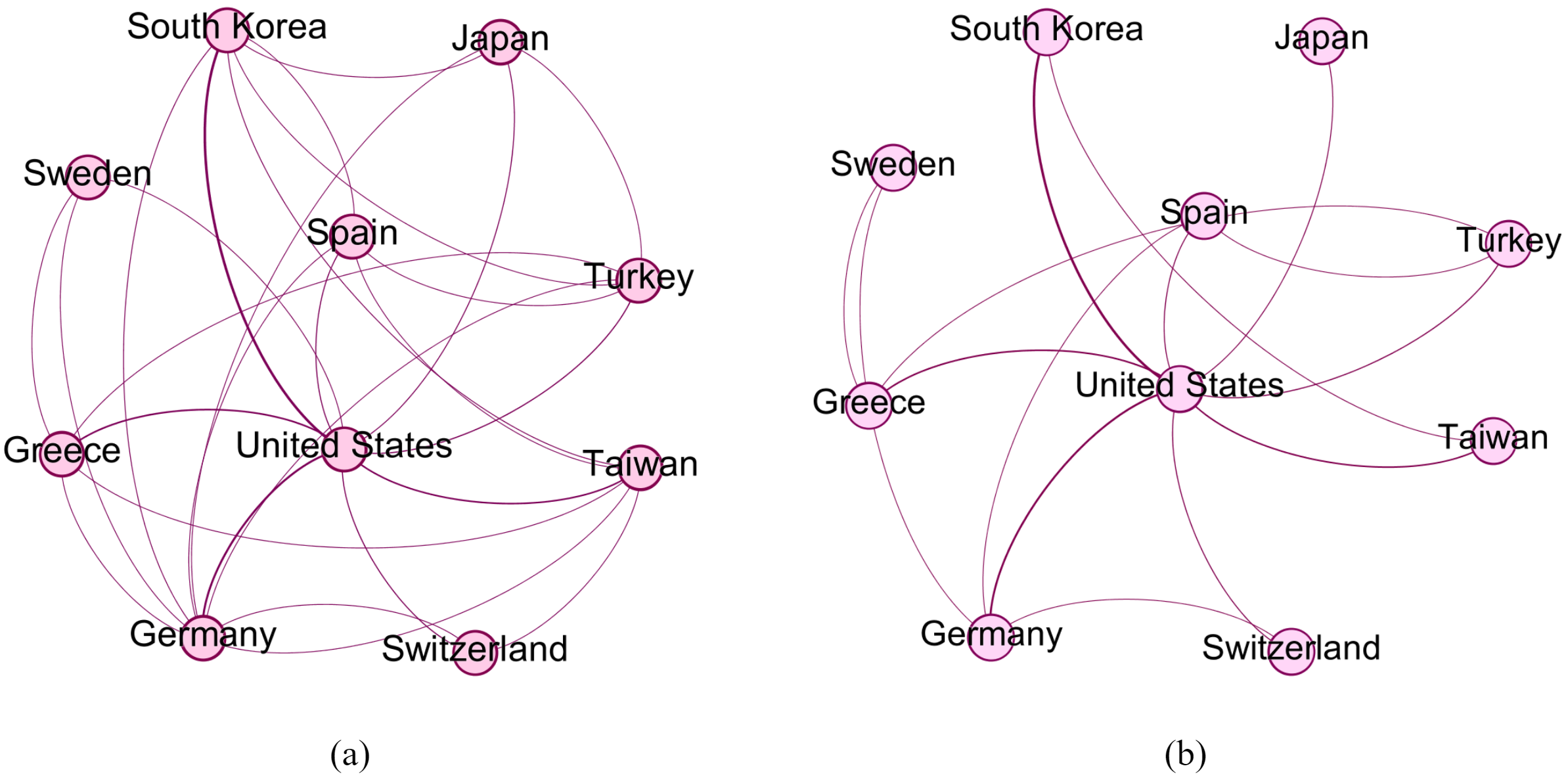}
\caption{A sampled subgraph of the advisor-student relationship network.}
\label{fig:fig1}
\end{figure}

The disparity filter is a graph sparsification algorithm utilized to effectively reduce the number of edges in a network while preserving its multi-scale nature~\cite{serrano2009extracting}. We apply this algorithm to remove insignificant edges from the advisor-student relationship network. Figure~\ref{fig:fig1} provides an example of the application of the disparity filter algorithm.

\FloatBarrier

\subsubsection*{Louvain community detection}
The Louvain community detection algorithm is utilized to identify communities within a large-scale network by optimizing the modularity. This algorithm aims to maximize the difference between the expected edge counts within a community and the actual edge counts. It employs a greedy approach with heuristics to solve the problem efficiently in polynomial time~\cite{Blondel_Guillaume_Lambiotte_Lefebvre_2008}. We apply this algorithm to detect communities within the advisor-student relationship network.

\subsubsection*{Leiden community detection}
The Leiden community detection algorithm is an advancement of the Louvain algorithm. It employs a fast local move approach and iteratively refines partitions to ensure the connectedness of all detected communities. Compared to the Louvain algorithm, it offers improved speed and provides more accurate partitions~\cite{Traag2019}. We use this algorithm to identify communities within the advisor-student relationship network.

\begin{figure}[!ht]
\centering
\includegraphics[width=\hsize, keepaspectratio]{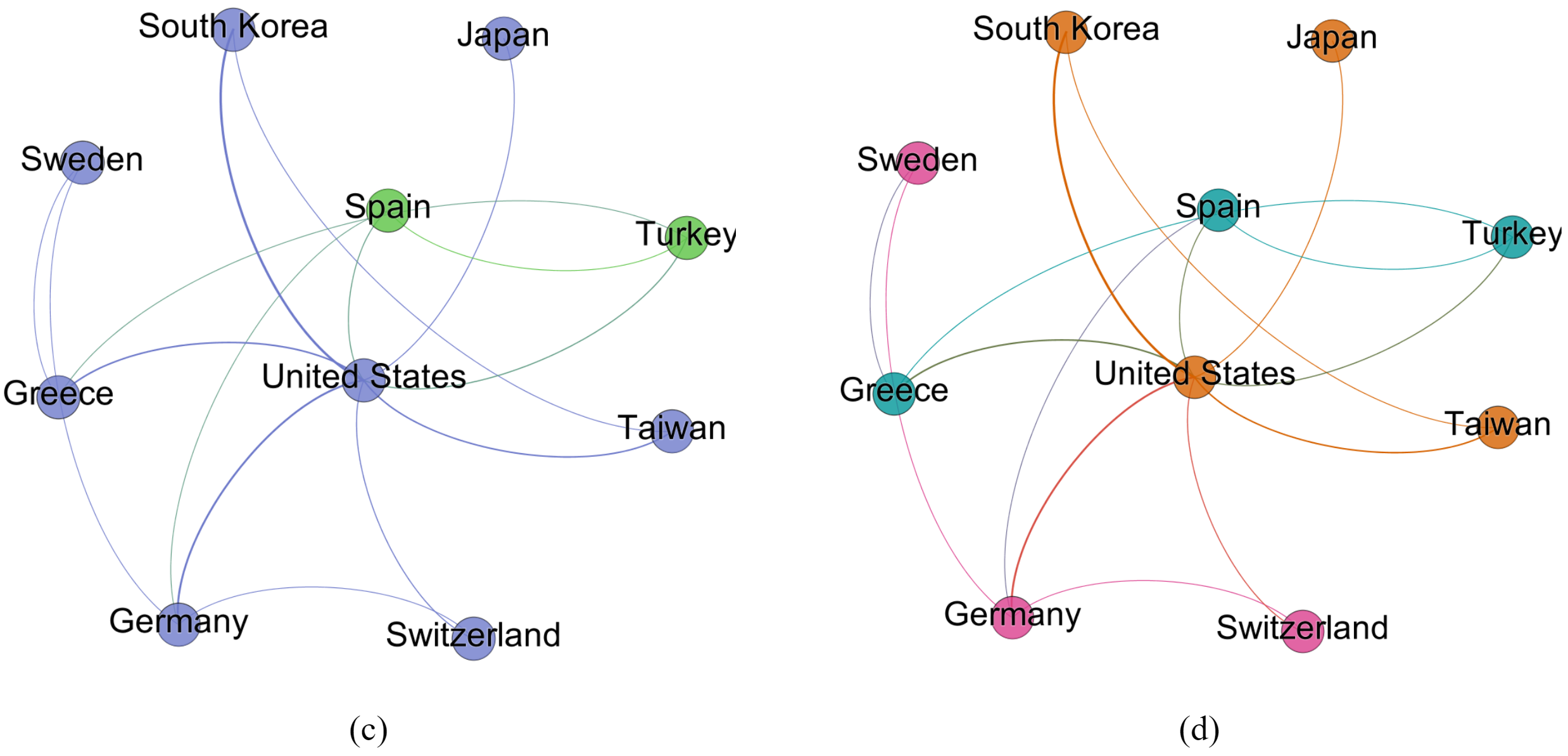}
\caption{The sampled subgraph of the advisor-student relationship network with specified communities.}
\label{fig:fig2}
\end{figure}

Figure~\ref{fig:fig2} shows the examples of the Louvain and Leiden community detection algorithms.

\FloatBarrier

\subsection*{Statistical analysis}
In this part, we provide a description of the statistical methods employed in this study.

\subsubsection*{Proportion hypothesis test}
The proportion hypothesis test is a statistical method that compares the ratio of an attribute in a population with a reference proportion. It also establishes a range of values that are likely to include the population proportion~\cite{diez2012openintro}. We utilize this technique to assess our research questions regarding biases in graduate admission.

\subsubsection*{Mann-Kendall test}
The Mann-Kendall test is a nonparametric method that assesses the presence and direction of trends. It is particularly suitable for detecting monotonic trends that exhibit consistent increases or decreases over time~\cite{Mann1945}. We employ this technique to evaluate the trends of variables such as gender/nationality diversity over time.

\subsection*{Metrics}
In this part, we provide the definitions of the measures that we calculate in this study. 

\subsubsection*{Weighted degree centrality}
Weighted degree centrality is defined for each node in a network by summing the weights of the edges connected to that node. The formula for weighted degree centrality is as follows:

\begin{eqnarray}
    WD(u) = \sum_{v}{w(v, u)},
\end{eqnarray}

where $v$ is a neighbor of $u$, and $w(v, u)$ is the weight of the edge between $v$ and $u$~\cite{Wei2012}. We employ this measure to examine the faculty members from which countries accept a greater number of students from other countries.

\subsubsection*{Closeness centrality}
For each node, closeness centrality is defined as the average distance between that node and all other nodes in the network. The formula for closeness centrality is as follows:

\begin{eqnarray}
    C(u) = \frac{n - 1}{\sum_{v=1}^{n-1} d(v, u)},
\end{eqnarray}

where $n$ represents the number of vertices that node $u$ is reachable from, and $d(v, u)$ denotes the geodesic distance between nodes $v$ and $u$~\cite{freeman1978centrality}. We utilize this metric to determine which countries are closer to the rest of the world in terms of admission results.

\subsubsection*{Entropy}
The entropy of a variable is defined as the average uncertainty of that variable based on its probability distribution. The formula for entropy is as follows:

\begin{eqnarray}
    E(X) = -\sum_{v=1}^{n} p_i \log p_i,
\end{eqnarray}

where the base of the logarithm is $e$, and $p_i$ represents the probability of the $i$-th outcome in variable $X$~\cite{renyi1959dimension}. We employ this measure to calculate the diversity of an advisor's research team.

\subsection*{Dataset}
Data collection was the most challenging aspect of this study. We collected data from multiple websites, each with its own unique structure, using a combination of manual and automated approaches.

The data collection procedure consists of four steps: manual data gathering, data collection using crawlers, removal of unnecessary data, and preprocessing. We collected data from the top 25 universities in North America, as ranked by Quacquarelli Symonds (QS) in 2021 for computer science \cite{topuniversitiesWorldUniversity}.

\subsubsection*{Manual data collection}
Among all the faculty members in the computer science departments of each university, we randomly selected approximately 30 professors. We collected information such as the professor's academic rank, home country, gender, research areas, and academic performance metrics (h-index and citation count). We also completed the prior universities (alma maters) column by referring to the professors' resumes and information available on their websites, LinkedIn, and Google Scholar. To determine the gender, we relied on images or pronouns specified on their websites. If the birthplace was not explicitly stated, we used the location of their undergraduate university to determine their home country. We also gathered academic records, such as citation counts and h-indexes, from Google Scholar.

The academic rank of faculty members, including Assistant Professor, Associate Professor, and Professor, was typically available on the university's website. Table~\ref{tab:tab1} presents the key information about faculty members that we collected from the university homepage and the professors' personal pages.

\begin{table}[ht]
\begin{adjustwidth}{-1cm}{}
\centering
\scriptsize
\caption{Essential professor information.}
\begin{tabular}{cccccccccc}
\toprule
University & ID & 
\begin{tabular}{@{}c@{}}Academic \\ Rank\end{tabular}
& Gender & 
\begin{tabular}{@{}c@{}}Home \\ Country\end{tabular}
& \begin{tabular}{@{}c@{}}Previous \\ Universities\end{tabular} &
\begin{tabular}{@{}c@{}}Citation \\ Count\end{tabular}
& h-index & 
\begin{tabular}{@{}c@{}}Publication \\ Count\end{tabular} &
\begin{tabular}{@{}c@{}}First Paper \\ Year\end{tabular}\\
\midrule
CMU & advisor37 & Professor & Female & United States & MIT/MIT/MIT & 34062 & 57 & 328 & 1979\\
\midrule
Stanford & advisor311 & 
\begin{tabular}{@{}c@{}}Associate \\ Professor\end{tabular}
& Male & Slovenia & \begin{tabular}{@{}c@{}}University of \\ Ljubljana/CMU\end{tabular} & 103427 & 125 & 477 & 1999\\
\bottomrule
\end{tabular}
\label{tab:tab1}
 \end{adjustwidth}
\end{table}
\FloatBarrier

For the professor's field column, we initially obtained the professor's research interests from their website, resume, or in some cases, from Google Scholar. Next, we manually determined whether the professor's research interests were associated with one or more of the 13 primary fields of the Association for Computing Machinery (ACM) computer science field category \cite{acmComputingClassification}. Table~\ref{tab:tab2} presents a sample mapping between professor interests and ACM subareas within our dataset.

\begin{table}[ht]
\centering
\scriptsize
\caption{Mapping between professor interests and ACM subareas.}
\begin{tabular}{ccc}
\toprule
ID & Field & Standard Field\\
\midrule
advisor96 & AI,ML,Optimization & Computing methodologies\\
\midrule
advisor445 & 
\begin{tabular}{@{}c@{}}cloud computing,databases,distributed \\ systems\end{tabular}
& 
\begin{tabular}{@{}c@{}}Software and its engineering/ \\ Information systems\end{tabular}\\
\midrule
advisor518 &
\begin{tabular}{@{}c@{}}natural language processing,computational \\ social sciences,machine learning\end{tabular}
&
\begin{tabular}{@{}c@{}}Computing methodologies/ \\ Information systems\end{tabular}\\
\bottomrule
\end{tabular}
\label{tab:tab2}
\end{table}
\FloatBarrier
After obtaining all the necessary information for each professor, we proceeded to gather the names of their students and any additional available information from their profiles. If any student-related information was available, we used it to populate the corresponding column; otherwise, we left it blank and planned to update it later with data collected from our crawlers in the next stage. Furthermore, after running the crawlers, we manually cross-checked the data to fill in any gaps using information available from other sources. The process of finding the information and collecting the data proved to be challenging and time-consuming, leading to the development of crawlers for different sections. Table~\ref{tab:tab3} presents the student information available in our dataset.

\begin{table}[ht]
\centering
\scriptsize
\caption{Student information in our dataset.}
\begin{tabular}{cccccc}
\toprule
ID & Degree & Start Year & Gender & Home Country & Previous Universities\\
\midrule
student2760 & MS & 2019 & Female & South Korea & Yonsei University\\
\midrule
student3575 & PhD & 2020 & Male & China & Tsinghua University\\
\midrule
student10241 & PhD & 2021 & Male & United States & Columbia University\\
\bottomrule
\end{tabular}
\label{tab:tab3}
\end{table}
\FloatBarrier

\subsubsection*{Data collection using crawlers}
We used the list of all students as input for the Google search engine to locate their websites and resumes, including their LinkedIn accounts. The next challenge was to automatically extract the required data from these websites and resumes to populate the information columns, such as degree, admission year, and alma maters. We also performed data cleaning on the output from the crawler and merged it with the primary dataset to ensure consistency and completeness.

We used the Name2GAN website \cite{qcriAcuaAudience} to label a person's gender, if it was not manually identified. We checked the results of this tool for 3000 previously labeled data. The results show that the gender detection tool has an accuracy higher than 90\%. We used manual labeling for cases that gender detection uncertainty was high to enrich the quality of our dataset.

\subsubsection*{Irrelevant data removal}
Since we recorded information about all students associated with each randomly-selected professor, including visiting students, undergraduates, postdocs, masters, and PhD students, it was important to filter out irrelevant data and include only graduate students for our analysis. The final version of the dataset was completed on August 2, 2022, and it consisted of a total of 13,936 graduate students.

\subsubsection*{Preprocessing}
The preprocessing stage consists of two phases:
\begin{enumerate}
\item Preparing the input for the crawlers.
\item Preparing the data for analysis.
\end{enumerate}

The most crucial component of the preprocessing stage was creating a consistent list of institutions that could be used for analysis and for the Google Maps crawler. We also double-checked the address results for each university to ensure that the mapping between university and address was unique. As mentioned earlier, we used these addresses to identify the students' countries of origin. In some cases, the home countries of students were improperly reported as a state rather than the country, and we corrected this during the preprocessing stage. Once the home country column was filled out, we standardized the names of the countries and prepared them for analysis. Additionally, the admission year column required cleaning, as there were specific irrational values that were quickly corrected.

Students' home country is determined based on explicit specifications, if available. If not explicitly specified, we first consider the country from which they earned an associate degree. If that information is not available, we use the location of their undergraduate university to determine their home country. Additionally, we utilized a crawler for the Google Maps API to search for the location of universities and schools, which provided us with the necessary addresses for further analysis.

\subsection*{Data exploration}
In this part, we present an overview of the key features of our dataset in order to gain insights into their distributions.

\subsubsection*{Advisors’ gender}
\begin{figure}[ht]
\centering
  \includegraphics[width=10cm,
  keepaspectratio]{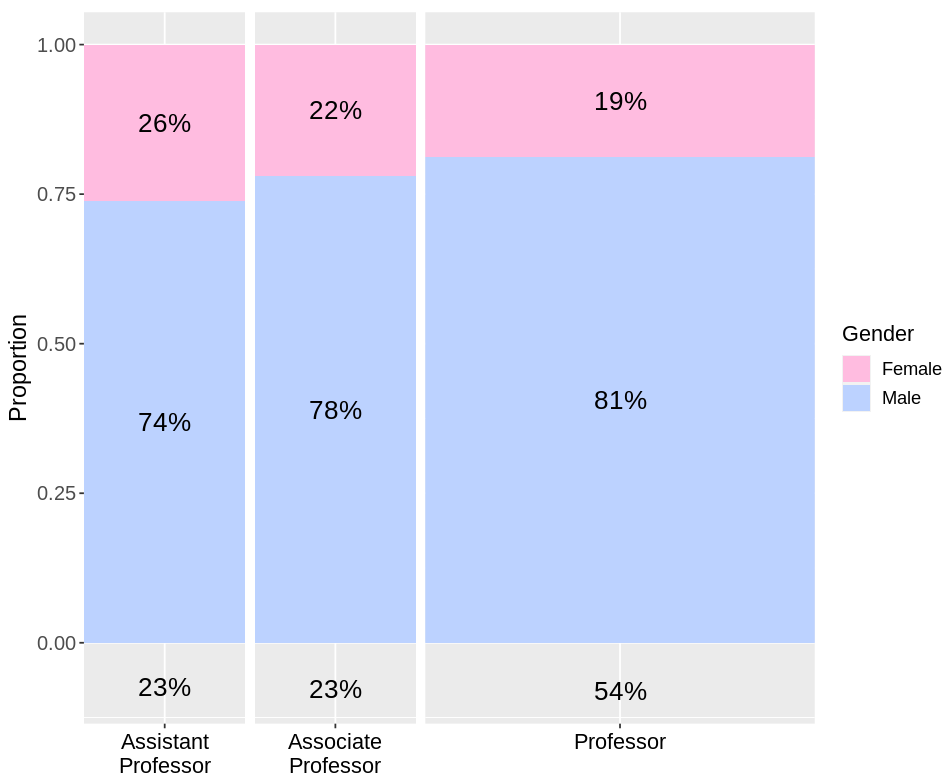}
\caption{Mosaic plot of advisors' academic rank and gender}
\label{fig:fig3}
\end{figure}

In this part, we examine the distributions of advisors' gender across other attributes. Figure~\ref{fig:fig3} displays the mosaic plot depicting the relationship between advisors' gender and their academic rank. The majority of advisors in our dataset hold the professor rank, and the highest proportion of male advisors is also observed at the professor level. This finding aligns with the results of~\cite{li2021gender}, which suggest that men have a greater likelihood of being promoted to the professor rank compared to women.

\FloatBarrier

\begin{figure}[h!]
\centerline{\includegraphics[width=\hsize, keepaspectratio]{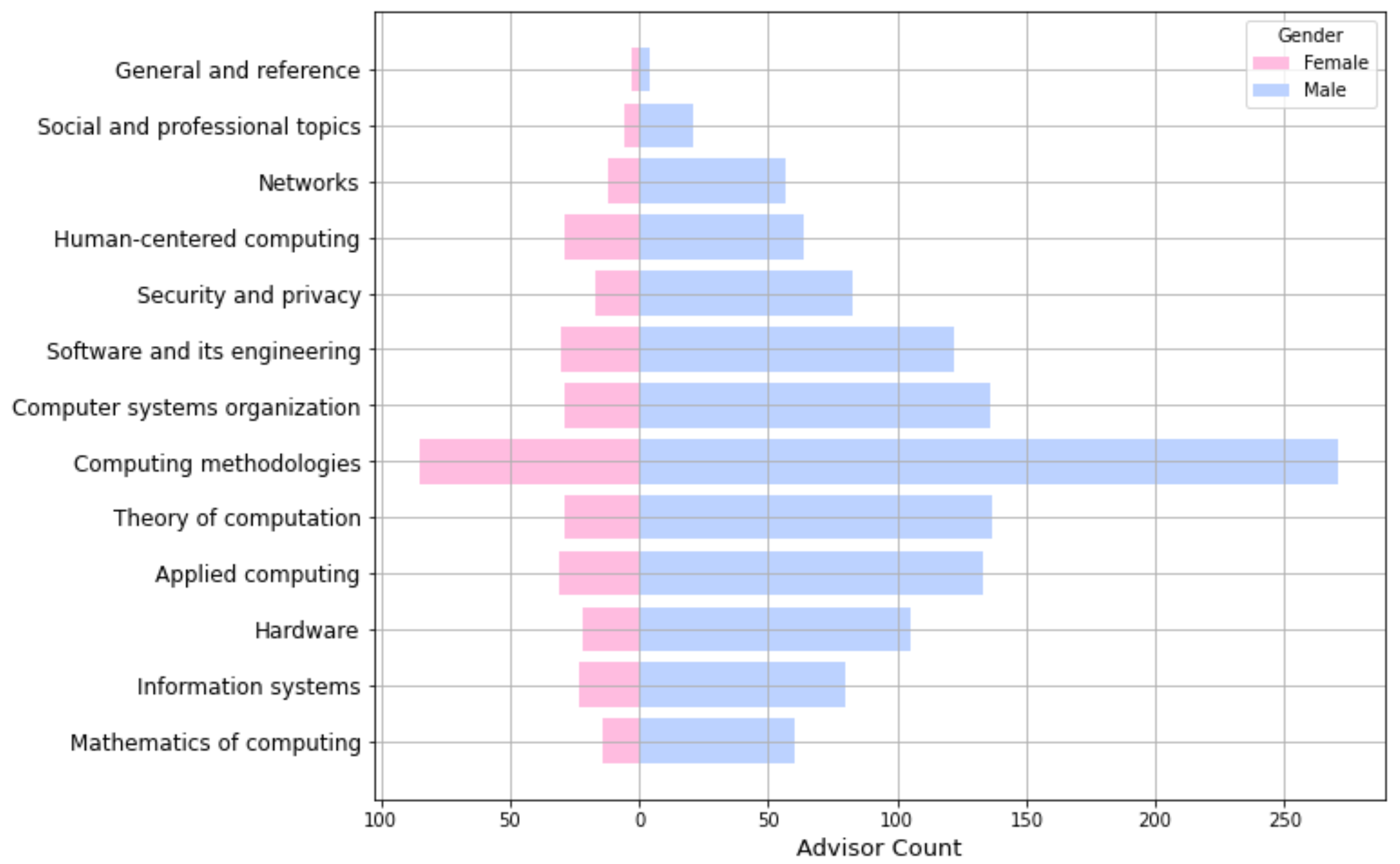}}
\caption{
  Back-to-back bar plot of advisors’ gender and their research fields.
}
\label{fig:fig4}
\end{figure}

Figure~\ref{fig:fig4} illustrates the distribution of female and male faculty members across different subfields of computer science in our dataset. The graph shows that the computing methodologies subfield has the highest number of advisors. This observation can be attributed to the growing significance of Artificial Intelligence, which falls under the computing methodologies category and is an interdisciplinary field~\cite{liu2018artificial, schonemann1985artificial}. The theory of computation and computer systems organization subfields represent the second and third largest groups, respectively.

\FloatBarrier

\begin{figure}[h!]
\centerline{\includegraphics[width=\hsize, keepaspectratio]{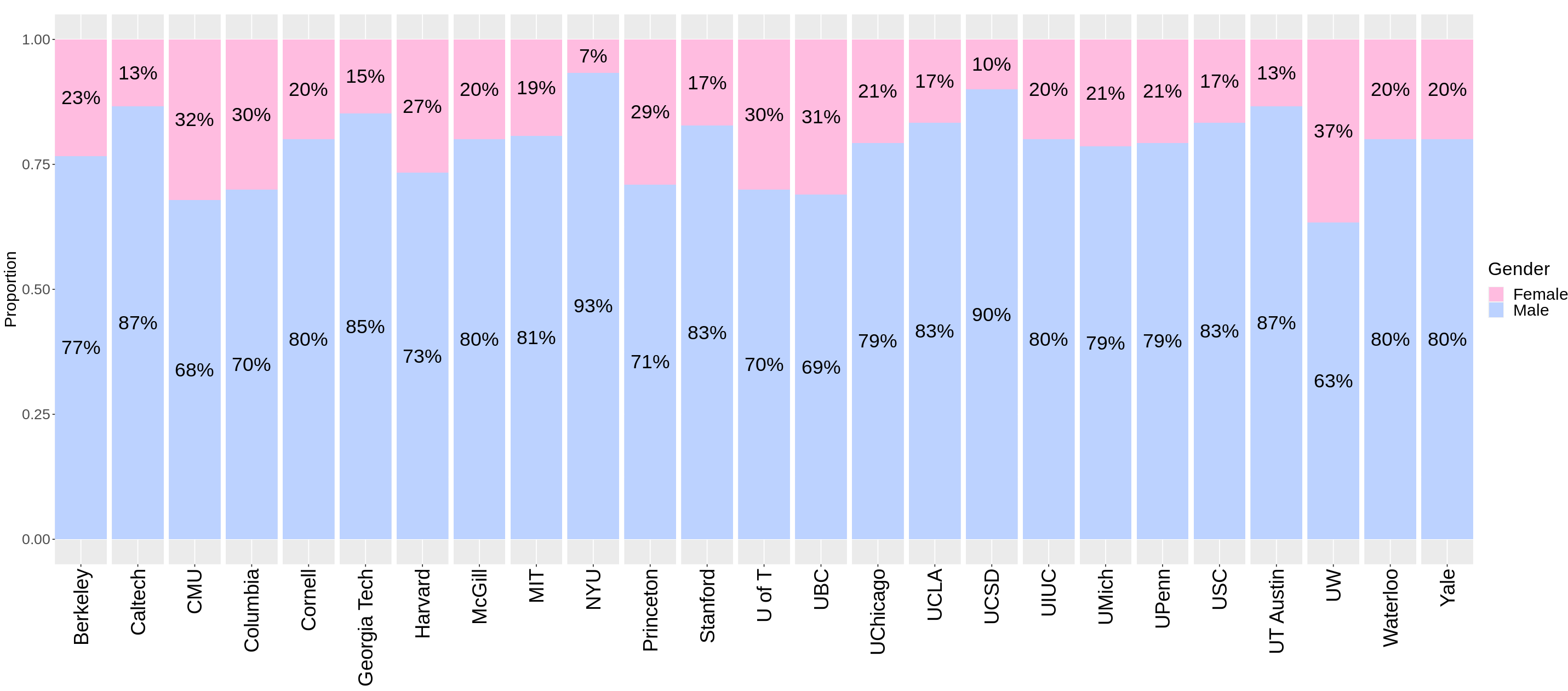}}
\caption{
  Gender-disaggregated bar plot showing the count of advisors for each university.
}
\label{fig:fig5}
\end{figure}

Figure~\ref{fig:fig5} shows the distribution of gender among computer science faculty members across different universities.

\FloatBarrier

\subsubsection*{Advisors’ academic performance metrics}
\begin{figure}[h!]
\centerline{\includegraphics[width=\hsize, keepaspectratio]{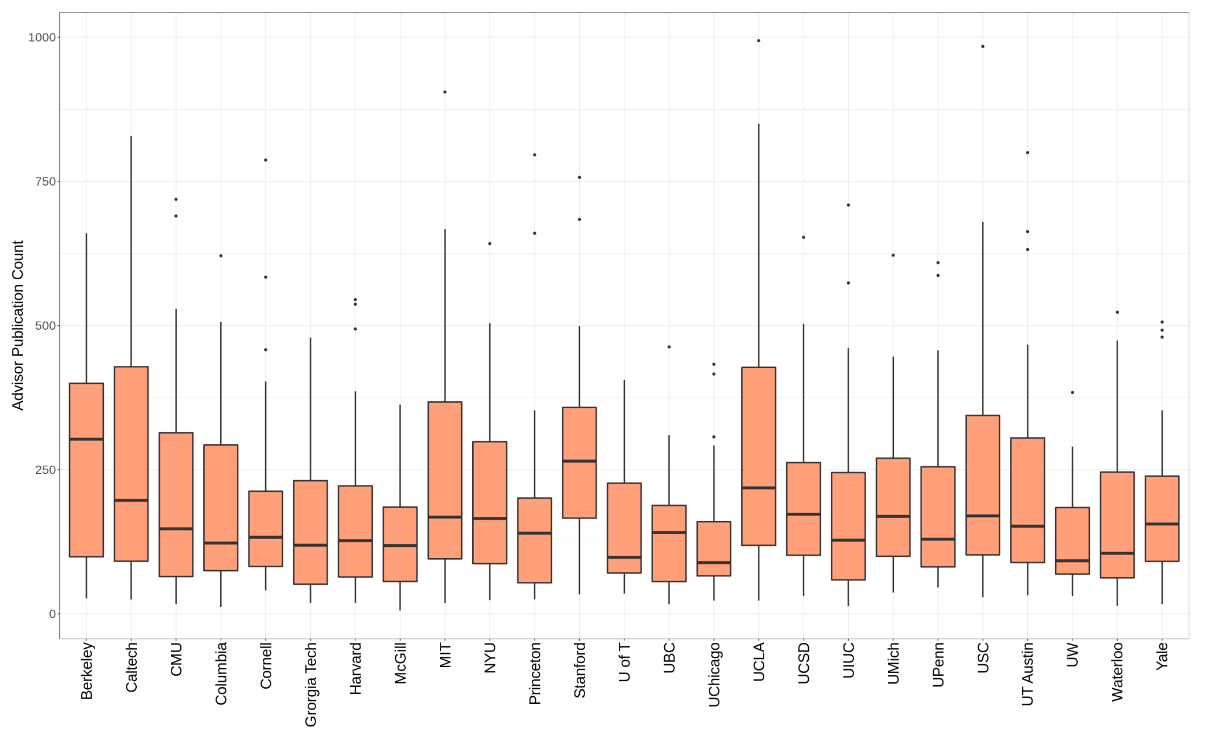}}
\caption{
  Boxplots of advisors’ publication counts for each university.
}
\label{fig:fig6}

\centerline{\includegraphics[width=\hsize, keepaspectratio]{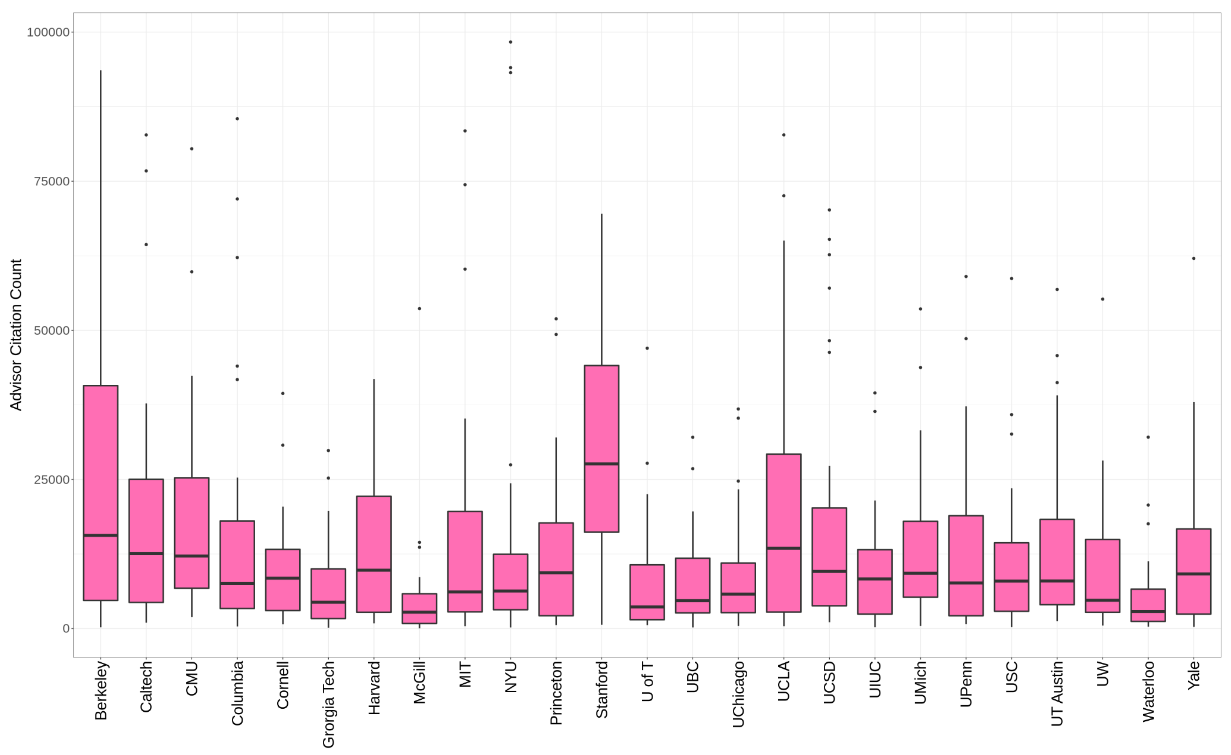}}
\caption{
  Boxplots of advisors’ citation counts for each university.
}
\label{fig:fig7}
\end{figure}

In this part, we present the dispersion of academic performance metrics of the faculty members, including publication count, h-index, and citation count, which are crucial indicators of the success of their research teams. Figure~\ref{fig:fig6} displays the boxplots of advisors' publication counts for each university. To enhance the resolution, advisors with more than 1000 publications were excluded from this plot.

Figure~\ref{fig:fig7} illustrates the distribution of citation counts for faculty members at each university. To improve the clarity of the diagram, faculty members with a citation count exceeding 100,000 were excluded.

\FloatBarrier

\begin{figure}[h!]
\centerline{\includegraphics[width=\hsize, keepaspectratio]{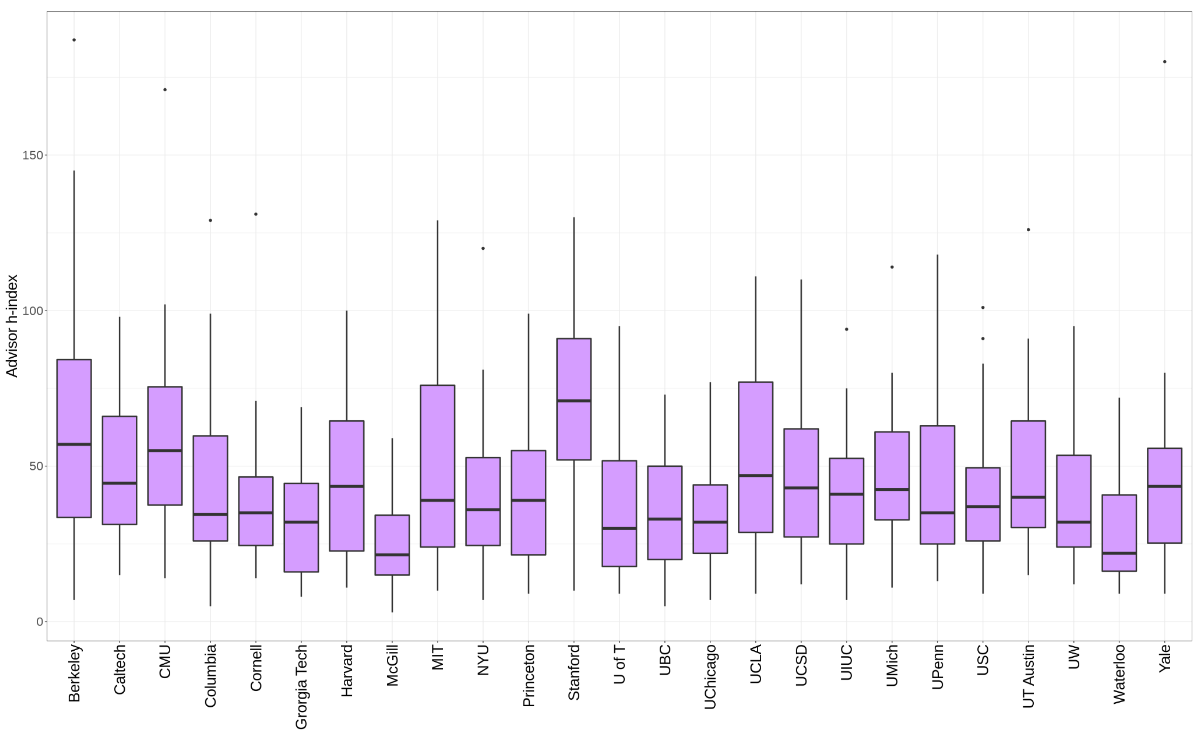}}
\caption{
  Boxplots of advisors’ h-indexes for each university.
}
\label{fig:fig8}
\end{figure}

Figure~\ref{fig:fig8} presents the boxplots of h-indexes for faculty members at each university. It is worth noting that the h-index metric has fewer outliers compared to the previous metrics, indicating that it may be a better indicator for assessing the success of research groups \cite{sharma2013h}.

\FloatBarrier

\subsubsection*{Students’ gender}
\begin{figure}[ht]
\centering
\includegraphics[width=9cm, keepaspectratio]{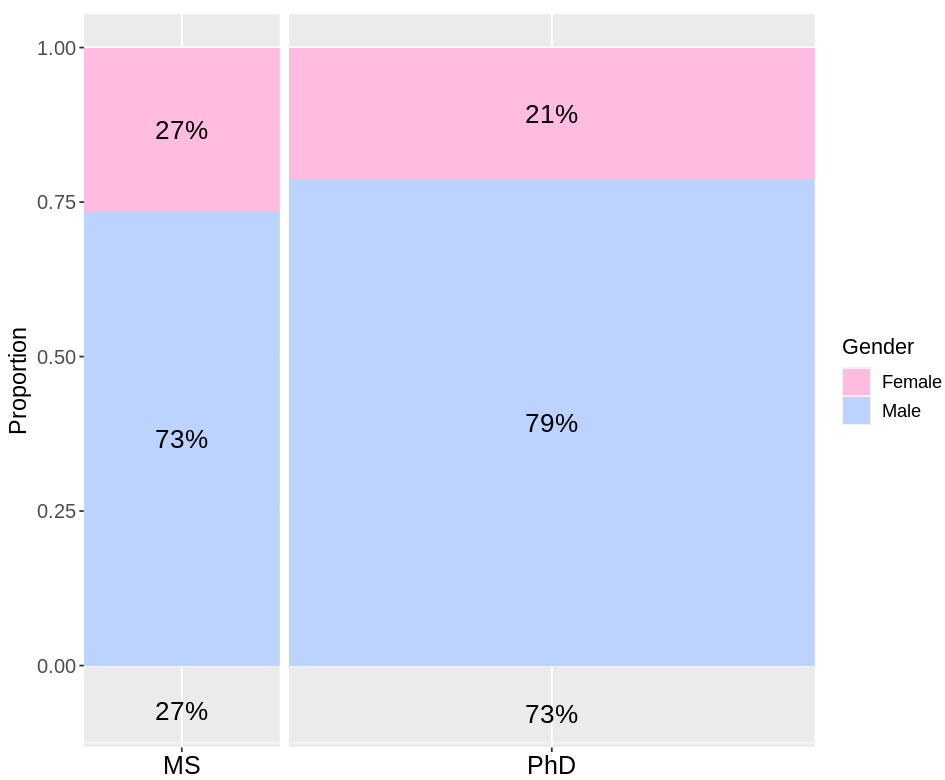}
\caption{Mosaic plot of students' degree and gender.}
\label{fig:fig9}
\end{figure}

In this part, we illustrate the distribution of students’ gender against other features. Figure~\ref{fig:fig9} presents a mosaic plot depicting the distribution of students' gender based on the degree they are pursuing (or have pursued) under the supervision of their advisor. The plot reveals that there are fewer women in graduate computer science programs, which aligns with the findings of Cuny and Aspray's study \cite{cuny2002recruitment}. Additionally, the female-to-male ratio decreases as the degree level progresses from masters to doctorate, potentially indicating a lower tendency among women to pursue higher education \cite{berg1983men}.

\FloatBarrier

\begin{figure}[h!]
\centerline{\includegraphics[width=\hsize, keepaspectratio]{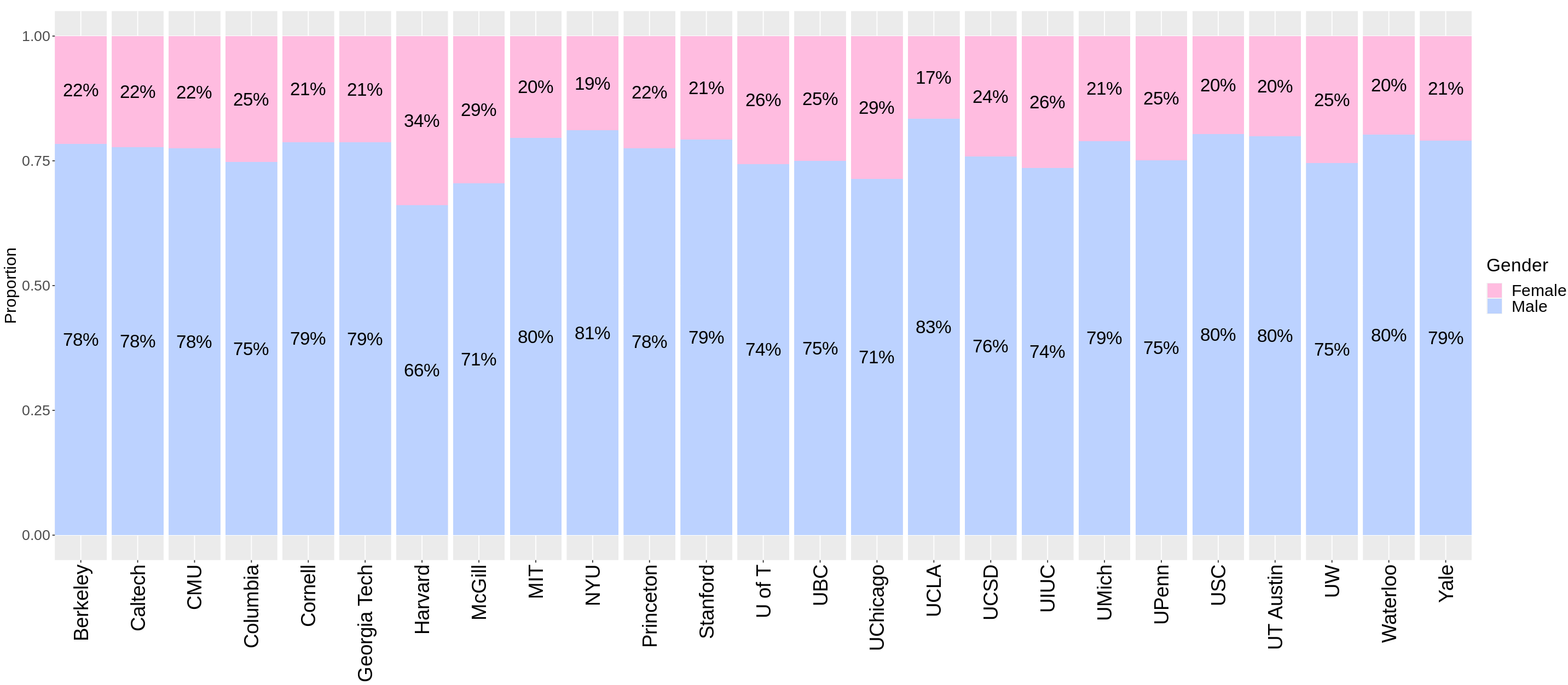}}
\caption{
  Gender-disaggregated bar plot showing the count of students for each university.
}
\label{fig:fig10}
\end{figure}

Figure~\ref{fig:fig10} displays the gender distribution of CS students across different universities.

\FloatBarrier

\subsubsection*{Nationality distributions}
\begin{figure}[!ht]
\centering
\includegraphics[width=9cm, keepaspectratio]{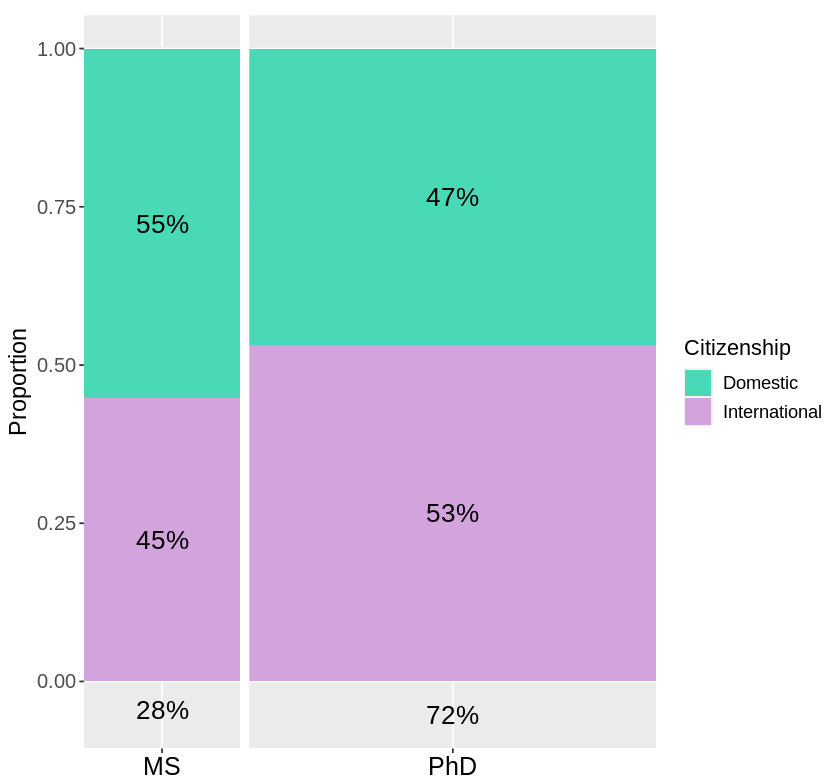}
\caption{Mosaic plot of students' degree and citizenship.}
\label{fig:fig11}
\end{figure}

In this part, we explore the distribution of nationalities among students and faculty members. Figure~\ref{fig:fig11} presents the distribution of students' citizenship for each degree. It shows that the majority of students apply for doctoral programs, and the percentage of international students is higher than that of American and Canadian students. This finding is in line with the result of a study by Okahana and Zhou~\cite{okahana2016graduate}, which states that in Fall 2015, approximately 55\% of students majoring in computer science or related programs were international students.

\FloatBarrier

\begin{figure}[!ht]
\centering
\includegraphics[width=\hsize, keepaspectratio]{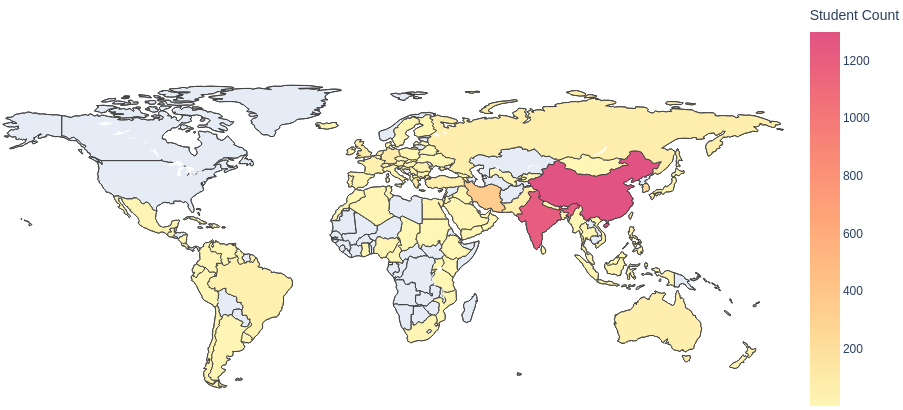}
\caption{Distribution of students’ nationalities.}
\label{fig:fig12}

\includegraphics[width=\hsize, keepaspectratio]{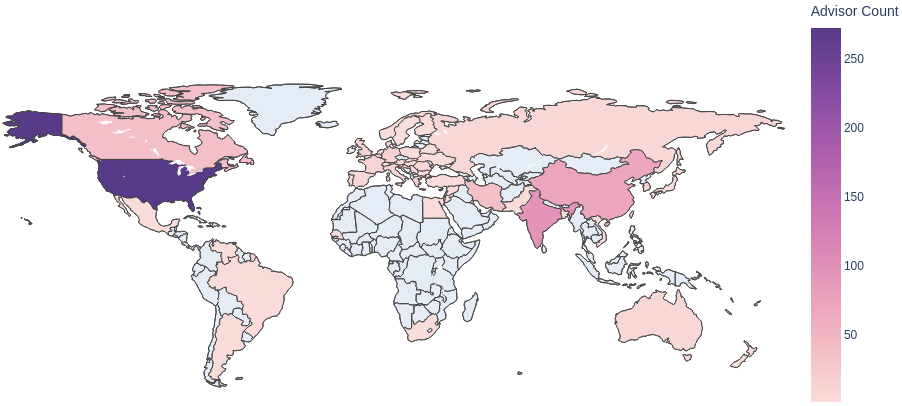}
\caption{Distribution of advisors’ countries of origin.}
\label{fig:fig13}
\end{figure}

Figure~\ref{fig:fig12} displays the distribution of students' nationalities on the world map. The United States and Canada have been excluded to focus solely on international students. The map reveals that the majority of international students are from China, India, and Iran, respectively. This finding aligns with the results of \cite{Sun2019}, which indicate that graduate programs are predominantly composed of Chinese and Indian students.

Figure~\ref{fig:fig13} displays the distribution of faculty members' home countries on the world map. The map reveals that the majority of advisors originated from the United States, followed by India, China, and Canada, respectively.

\FloatBarrier

\begin{figure}[h!]
\centerline{\includegraphics[width=\hsize, keepaspectratio]{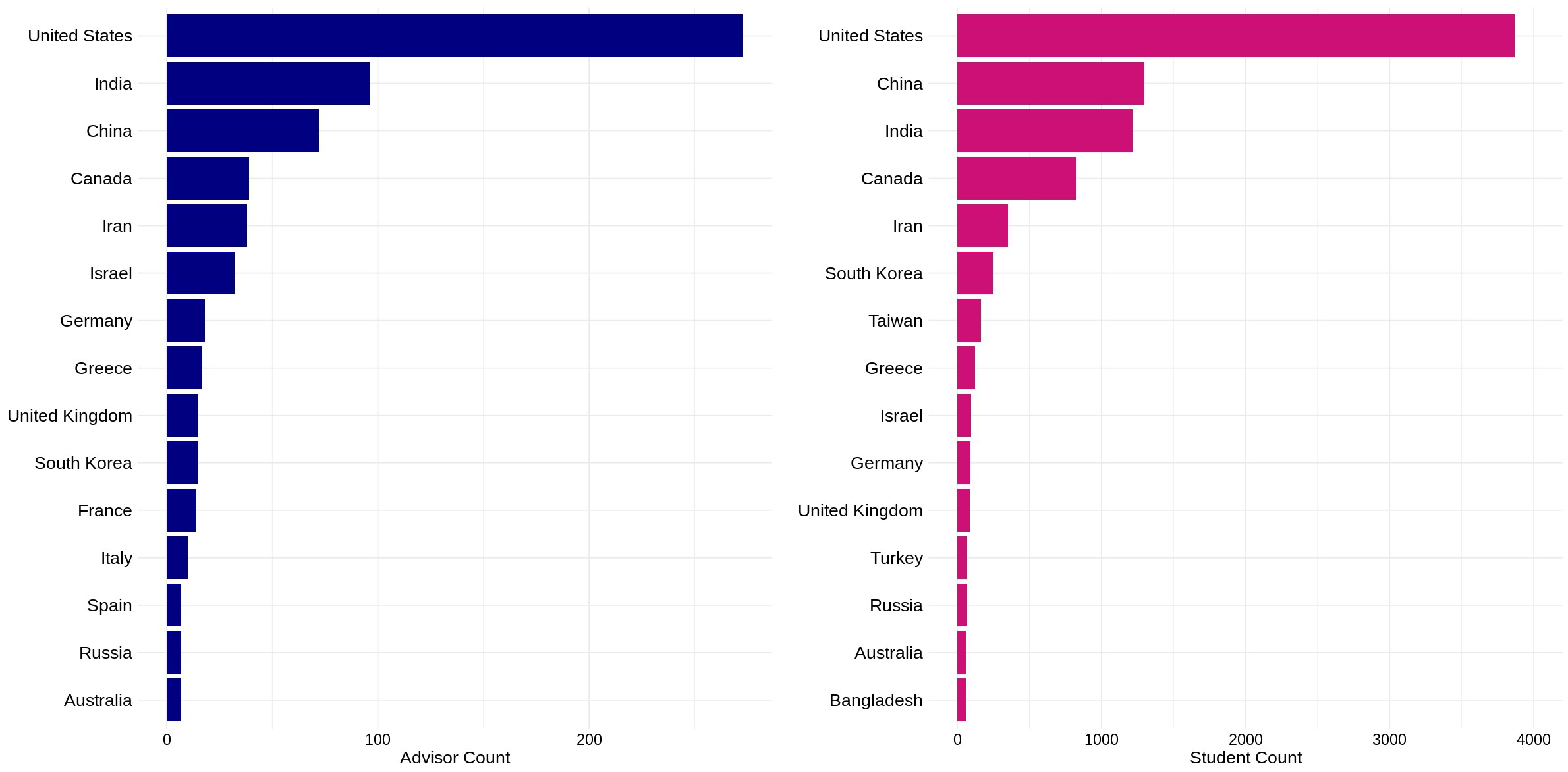}}
\caption{
  Distributions of countries of origin among faculty members and students.
}
\label{fig:fig14}
\end{figure}

Figure~\ref{fig:fig14} depicts the sorted bar plots of the 15 most common countries among faculty members and students.

\FloatBarrier

\section*{Results and discussion}
In this section, we provide a comprehensive explanation of our analyses and interpret the results we obtained.

\subsection*{Assessing gender partiality}
\begin{figure}[!ht]
\centering
\includegraphics[width=10cm, keepaspectratio]{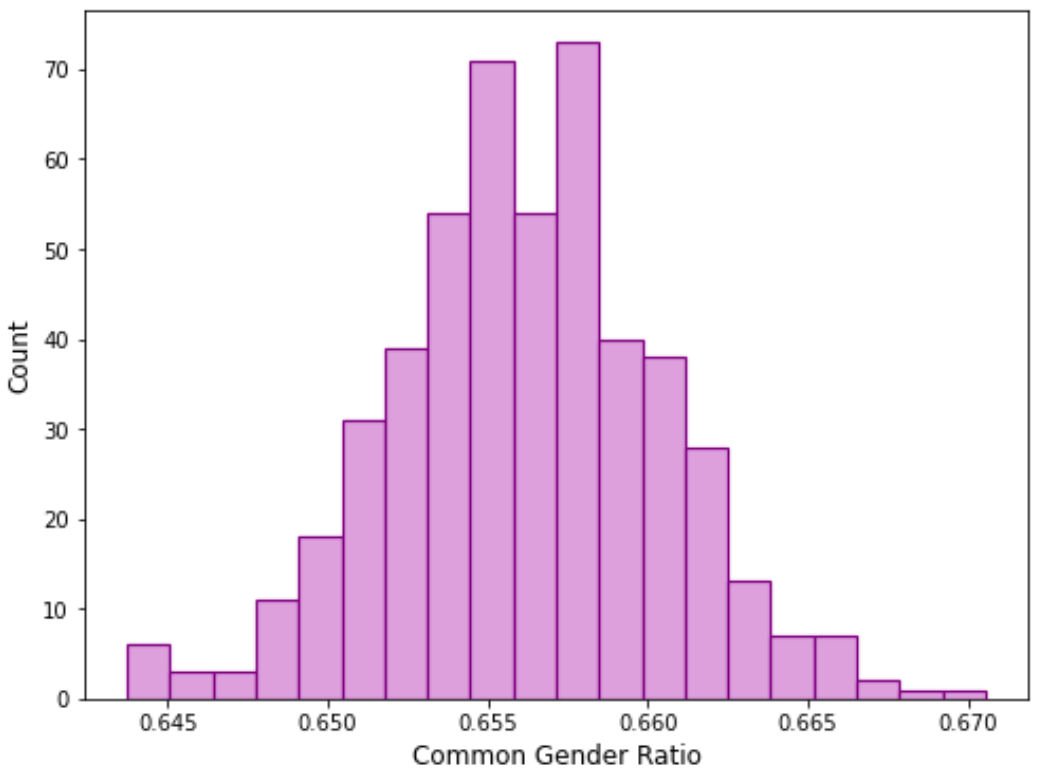}
\caption{Histogram of advisor-student common gender proportion.}
\label{fig:fig15}
\end{figure}

In this part, we evaluate the presence of gender bias in admission decisions. We conduct a two-sided hypothesis test with a significance level of 0.05 to examine whether there is gender bias in the acceptance of graduate students into advisors' research groups. To accomplish this, we employ a simulation-based approach with 500 iterations~\cite{diez2012openintro}. In each iteration, we generate 13,759 advisor-student pairs, where the gender of each component is selected based on the observed ratio in our dataset. Specifically, the probability of an advisor being male is 0.788, and the probability of a student being male is 0.771. This simulation yields an approximately normal distribution with a mean of 0.6562 and a standard deviation of 0.0212, as depicted in Figure~\ref{fig:fig15}. It is important to note that this distribution represents the values for the ratio of advisor-student pairs with the same gender, assuming no gender bias in admitting graduate students. In our dataset, the observed ratio of advisor-student pairs with the same gender is 0.6896. We will now test whether this observed value is likely to occur in the simulated distribution. Thus, our hypothesis test is formulated as follows:

\begin{eqnarray}
    \begin{gathered}
        H_0: p_{common \: gender \: ratio} = 0.6562 \\
        H_a: p_{common \: gender \: ratio} \neq 0.6562
    \end{gathered}
\end{eqnarray}

Using a z-test, we obtained a p-value of 0.1152, which is higher than the significance level of 0.05. Therefore, we cannot reject the null hypothesis. In other words, the data does not provide strong evidence of gender bias in the admissions of graduate students. This finding is consistent with the results of Maxwell's study~\cite{Maxwell1976}, which also concluded that gender is not a significant factor in graduate student acceptance.

\subsection*{Evaluating nationality bias}
\begin{figure}[!ht]
\centering
\includegraphics[width=10cm, keepaspectratio]{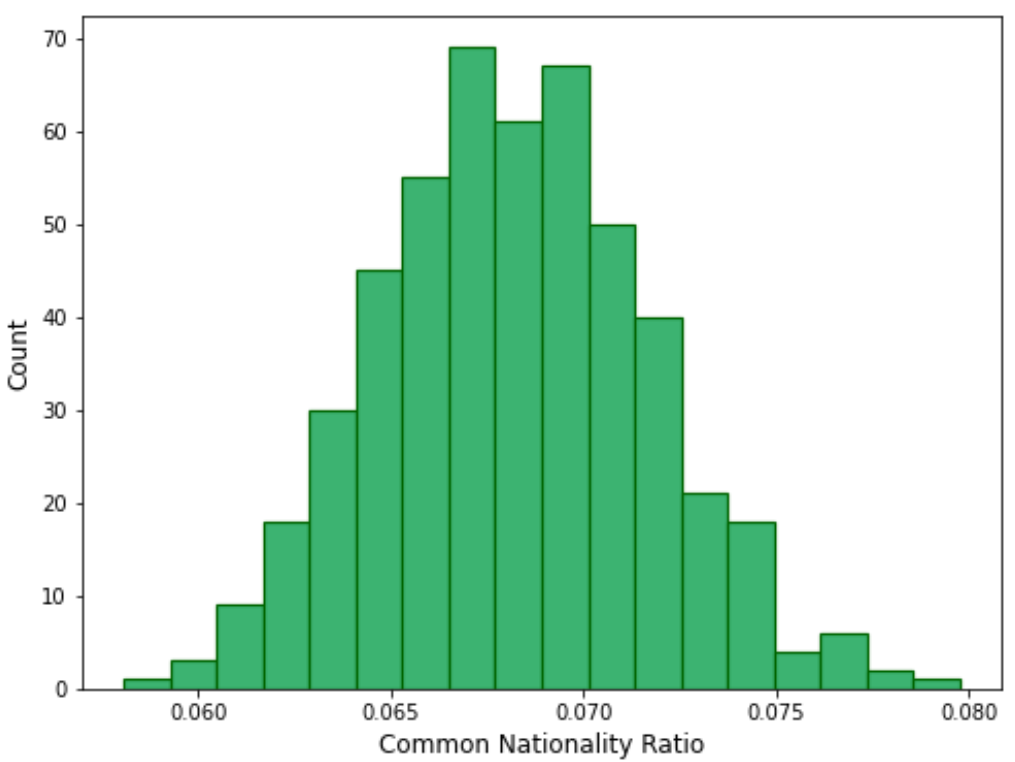}
\caption{Histogram of advisor-student common nationality ratio.}
\label{fig:fig16}
\end{figure}

In this part, we aim to investigate the presence of nationality bias in advisor-student relationships. We conduct a two-sided hypothesis test with a significance level of 0.05 to assess the existence of such bias. Similar to the previous analysis, we employ a simulation-based approach with 500 iterations.
For this analysis, we only consider international students who are not from the United States or Canada. At each iteration, we generate 4,839 advisor-student pairs, where the nationality of each individual is selected with a probability equal to the observed ratio in the dataset. In each iteration, we calculate the ratio of advisor-student pairs with the same nationality. The resulting distribution, shown in Figure~\ref{fig:fig16}, approximates a normal distribution with a mean of 0.0682 and a standard deviation of 0.0113.
In our dataset, the proportion of advisor-student pairs with the same nationality is 0.1593. To assess the likelihood of observing such a ratio in the simulated distribution, we formulate the following hypothesis test:

\begin{eqnarray}
    \begin{gathered}
        H_0: p_{common \: nationality \: ratio} = 0.0682 \\
        H_a: p_{common \: nationality \: ratio} \neq 0.0682
    \end{gathered}
\end{eqnarray}

Using a z-test, we obtain a p-value of $p < 10^{-15}$, which is significantly lower than the chosen significance level of 0.05. Therefore, we reject the null hypothesis and conclude that there is strong evidence of nationality bias in admitting international graduate students. This bias may be attributed to advisors' familiarity with universities in their home country and their potential to make more accurate assessments of students who have graduated from those universities.

\subsection*{Advisor-student relationship network}
\begin{figure}[ht]
\centering
\includegraphics[width=\hsize, keepaspectratio]{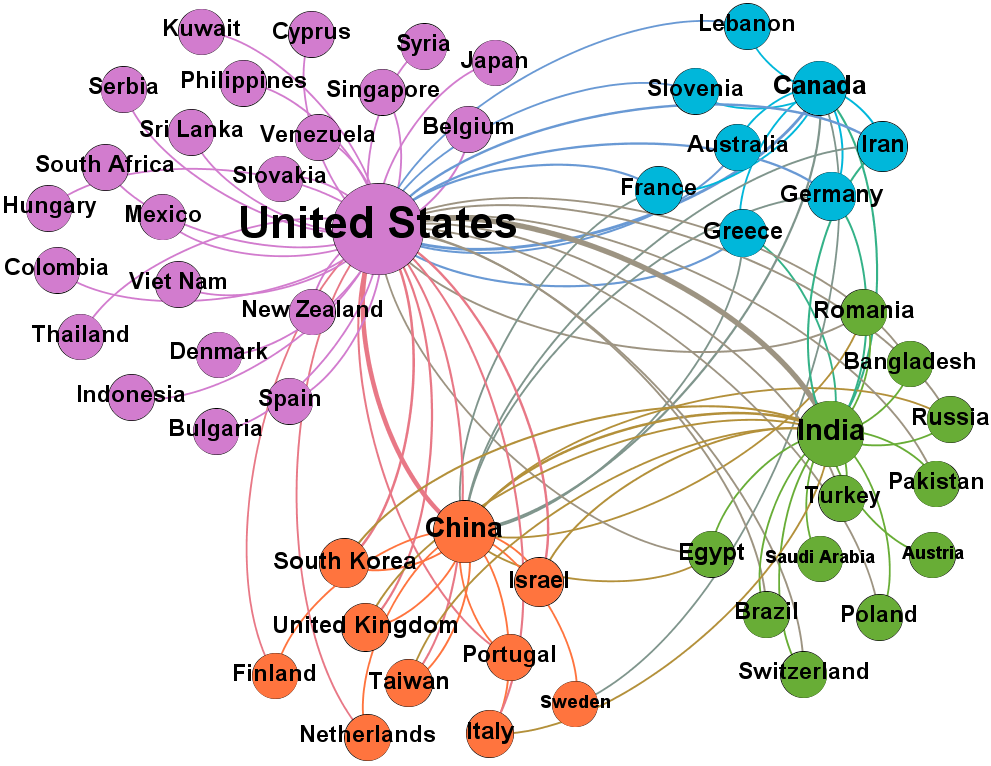}
\caption{Cross-country advisor-student relationship network, with communities detected via Louvain algorithm.}
\label{fig:fig17}

\includegraphics[width=\hsize, keepaspectratio]{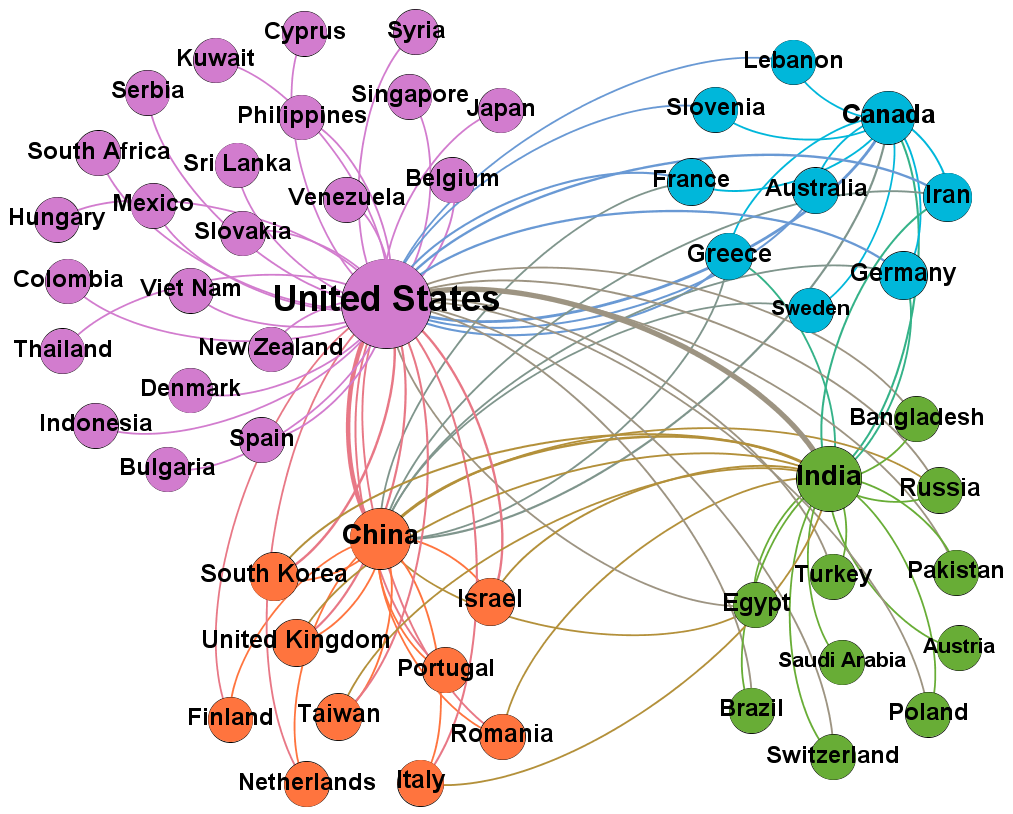}
\caption{Cross-country advisor-student relationship network, with communities detected via Leiden algorithm.}
\label{fig:fig18}
\end{figure}

In this part, we present a cross-country advisor-student relationship network based on our dataset. The network is constructed by connecting the nationalities of students and their advisors with weighted edges. We apply the disparity filter algorithm~\cite{serrano2009extracting} to eliminate insignificant edges and remove isolated nodes from the network. Figure~\ref{fig:fig17} provides an overview of the advisor-student relationship network. In this visualization, the size of the nodes and labels corresponds to the weighted degree and closeness centralities, respectively. The thickness of the edges represents their weight, which indicates the number of advisor-student pairs between the respective countries. Additionally, the nodes are color-coded based on their community assignment, determined using the Louvain community detection algorithm~\cite{Blondel_Guillaume_Lambiotte_Lefebvre_2008}.

The countries with the highest values for both centrality metrics are the United States, India, China, Canada, and Iran, respectively. This observation aligns with the previous findings that faculty members from these countries are prevalent in top universities. It serves as further evidence of the potential existence of nationality bias in advisor-student relationships.

In Figure~\ref{fig:fig18}, the advisor-student relationship network is depicted with similar settings, but the Leiden algorithm~\cite{Traag2019} is utilized for community detection. According to the results, Sweden and Romania are assigned to different communities compared to Figure~\ref{fig:fig17}.

\FloatBarrier

\subsection*{Exploring time effect}
\begin{figure}[!ht]
\centering
\includegraphics[width=\hsize, keepaspectratio]{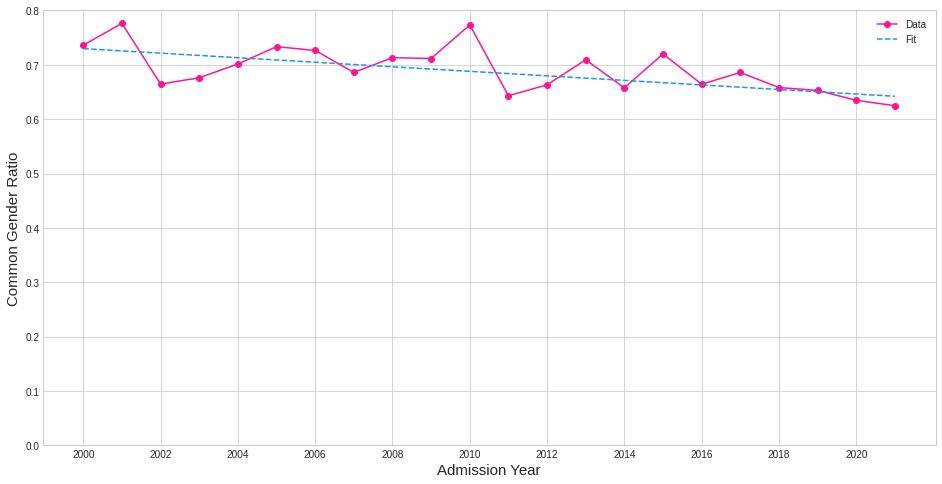}
\caption{Time series of advisor-student identical gender ratio.}
\label{fig:fig19}

\includegraphics[width=\hsize, keepaspectratio]{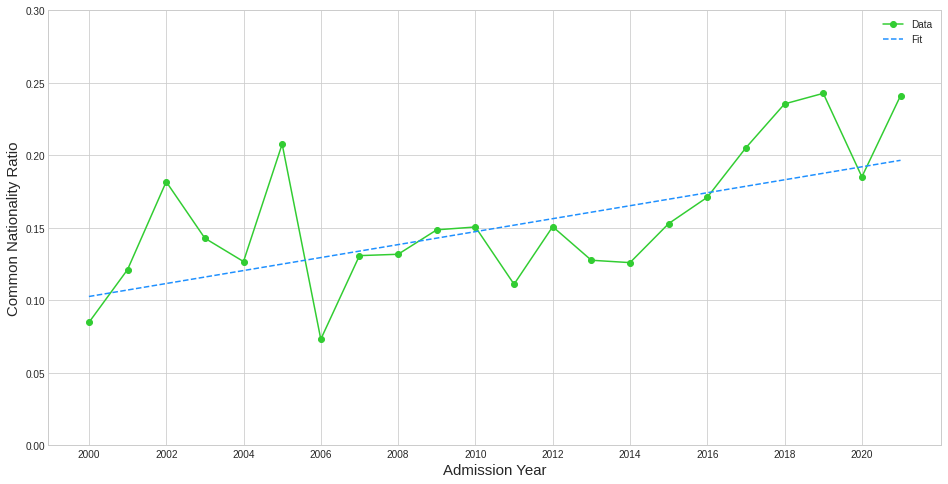}
\caption{Time series of advisor-student similar nationality ratio.}
\label{fig:fig20}
\end{figure}

In this part, we analyze the changes in bias patterns over time. Specifically, we examine admissions from 2000 to 2021. For each year, we calculate the ratios of advisor-student pairs with the same gender and nationality. Figure~\ref{fig:fig19} illustrates the time series of the identical gender ratio. The results of a Mann-Kendall test indicate that this time series exhibits a statistically significant decreasing trend ($p < 0.01$).

Figure~\ref{fig:fig20} illustrates the proportions of advisor-student pairs with the same nationality across different acceptance years. We observe an increasing trend in these proportions, which is consistent with the results of a Mann-Kendall test ($p < 0.01$).

\subsection*{Investigating relationship between academic success and diversity}
In this part, we aim to investigate whether there is a correlation between diversity in advisors' research groups and their academic success. To assess this relationship, we employ scientometrics, which are described in Table~\ref{tab:tab4}, as measures of research group success.

\begin{figure}[!ht]
\centering
\includegraphics[width=\hsize, keepaspectratio]{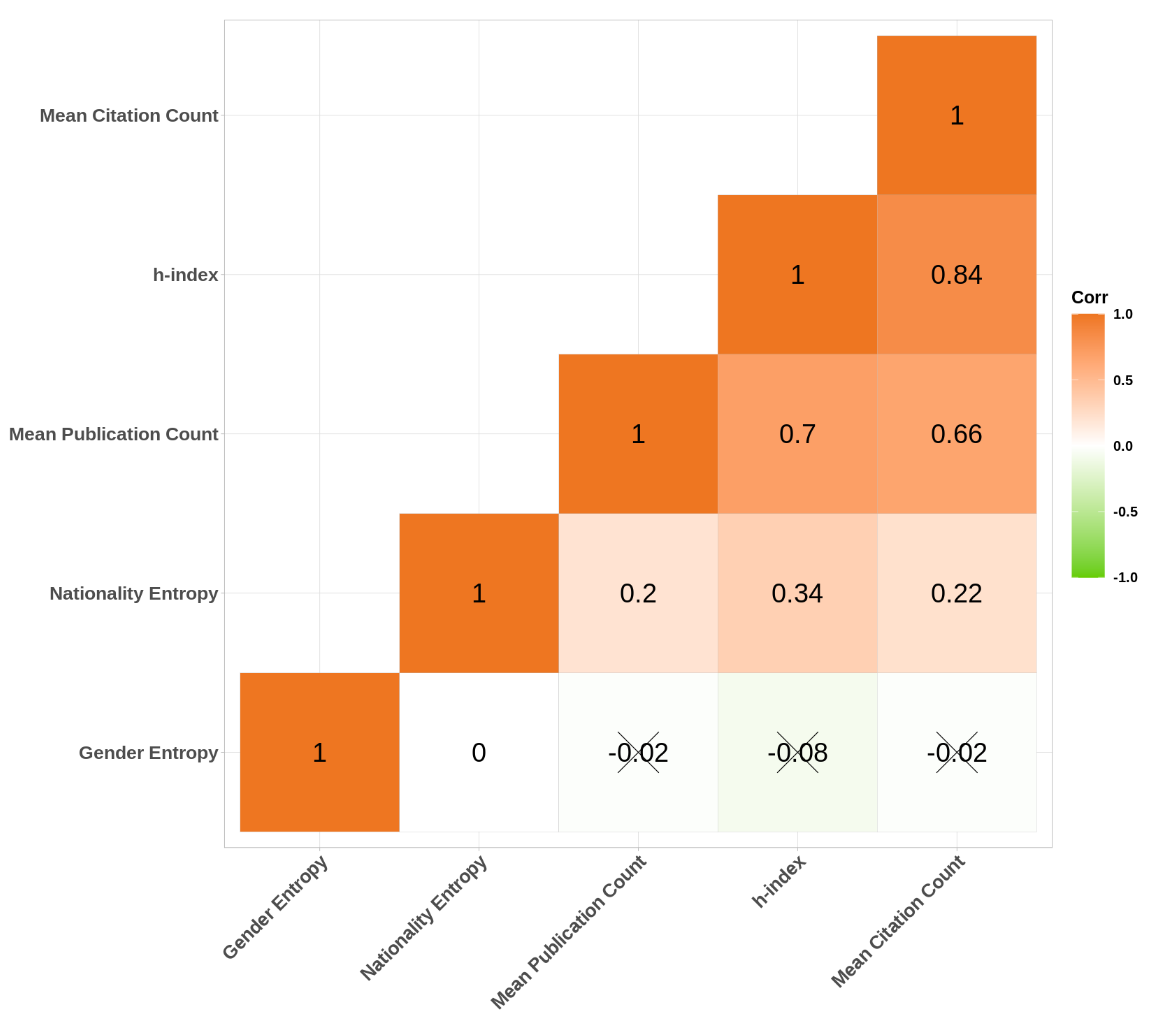}
\caption{Correlogram of academic success measures and gender/nationality diversity.}
\label{fig:fig21}
\end{figure}

\begin{table}[ht]
\centering
\scriptsize
\caption{Scientometrics and their explanations.}
\begin{tabular}{cc}
\toprule
Metric & Explanation\\
\midrule
h-index & Advisor’s h-index\\
\midrule
Mean Citation Count & 
\begin{tabular}{@{}c@{}}Advisor’s number of citations divided by \\ the years of her/his presence in academia\end{tabular}\\
\midrule
Mean Publication Count & 
\begin{tabular}{@{}c@{}}Advisor’s number of publications divided by \\ the years of her/his research experience\end{tabular}\\
\bottomrule
\end{tabular}
\label{tab:tab4}
\end{table}
\FloatBarrier
Moreover, we consider the entropy of genders and nationalities among an advisor's students as measures of diversity within their research group. We calculate the academic success and diversity measures for 737 advisors in our dataset. Subsequently, we compute the correlations between these variables, as shown in Figure~\ref{fig:fig21}. To assess the statistical significance of each correlation, we conduct a hypothesis test with a significance level of 0.01.
Based on the results, the correlations between gender entropy and other variables are close to zero and not statistically significant. This suggests that there is no significant linear correlation between gender diversity and the performance of research groups. On the other hand, nationality entropy exhibits a moderate positive correlation with advisors' h-index. This implies that research teams with greater diversity in terms of nationality tend to have higher research productivity. Additionally, there are weak positive linear relationships between nationality diversity and the remaining academic success metrics. It is important to note that the h-index is considered a more reliable measure of academic success~\cite{sharma2013h}.

\FloatBarrier

\subsection*{Analyzing trends of diversity}
\begin{figure}[!ht]
\centering
\includegraphics[width=\hsize, keepaspectratio]{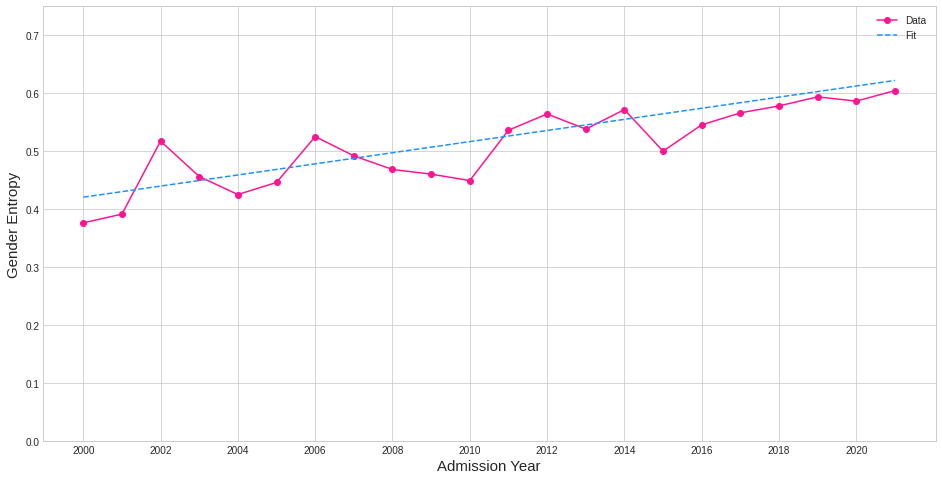}
\caption{Students’ gender entropy across admission years.}
\label{fig:fig22}

\includegraphics[width=\hsize, keepaspectratio]{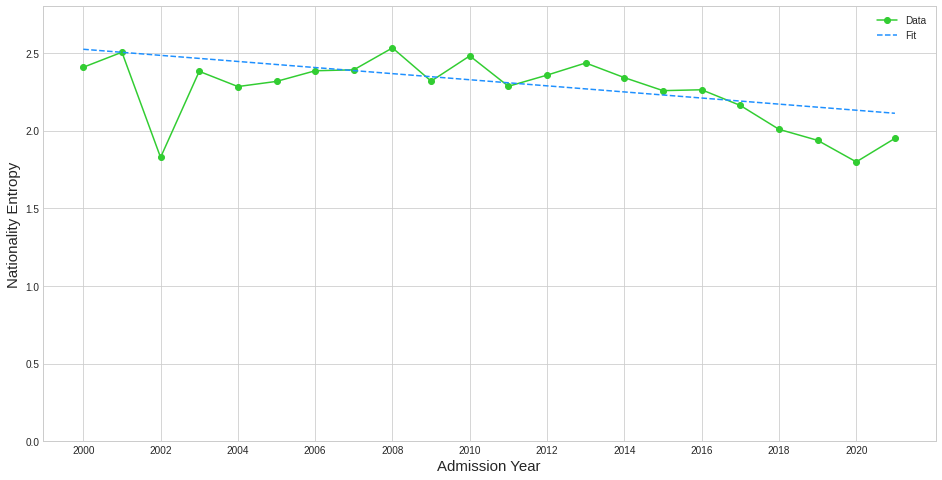}
\caption{Students’ nationality entropy across admission years.}
\label{fig:fig23}
\end{figure}

In this section, we discuss how gender and nationality diversities have changed over the past two decades. Once again, we employ the Mann-Kendall test to assess the strength of the observed trend. Figure~\ref{fig:fig22} illustrates the increasing trend in gender entropy over time. According to the results of the Mann-Kendall test, the observed trend is highly statistically significant ($p < 10^{-5}$).

Figure~\ref{fig:fig23} shows the time series of nationality entropy. As depicted, there has been a decrease in nationality diversity over time. The decline from 2016 to 2020 is particularly noticeable. The results of the Mann-Kendall test confirm that the observed trend is statistically significant ($p< 0.01$).

\FloatBarrier

\section*{Future work}
While our work presents a novel study analyzing gender and nationality biases in graduate admissions over recent decades, future research should aim to explore other crucial factors influencing admission decisions. These factors include academic background, religion, and politics, in order to provide a more comprehensive understanding of bias in graduate admissions. To achieve this, researchers could contemplate integrating our dataset with additional sources, such as institutional reports and the social media profiles of students and faculty members on platforms like Twitter, to glean fresh insights on this matter.

Moreover, another promising avenue for future research involves evaluating whether specific stages of the admissions process accentuate gender and nationality biases, and how these biases manifest diversely across various universities. For example, researchers could concentrate on distinct phases of the admissions process, such as committee decisions, to discern differing bias patterns.

Additionally, future investigations might delve into the correlation between gender and nationality diversity within computer science faculty and observed biases in graduate admissions. This analysis could yield insights into potential strategies for addressing these biases effectively.

Lastly, a valuable topic for future research could be assessing whether significant variations in gender and nationality biases exist across different subfields within computer science (e.g., artificial intelligence, systems, theory). Furthermore, exploring how these biases correlate with broader trends could provide valuable insights into the dynamics of bias within the field.

\section*{Conclusion}
In this study, we analyzed the distribution of genders and nationalities among students and their advisors. We conducted two-sided hypothesis tests to examine the presence of bias in gender and home country within advisor-student relationships. Our findings indicate that there is no gender bias in admission results. However, our results confirm the existence of bias against international applicants based on nationality. Additionally, we explored centrality metrics in the advisor-student relationship network, revealing that the United States, India, and China are the dominant countries in CS academia, influencing the composition of students and faculty members in top North American universities. We investigated the trends in gender and nationality bias over time and observed a reduction in gender bias, while nationality bias has shown an increasing pattern. Furthermore, we established a positive relationship between diversity in the nationalities of research group members and their academic performance. Lastly, we demonstrated an increase in gender diversity over time, alongside a decline in nationality diversity.

We acknowledge a limitation regarding the data collected for this study. We cannot guarantee that each faculty member consistently includes all individuals on their webpage. While the majority of computer science professors at high-ranking universities update their homepage at least once a year, some faculty members may not update information about newly admitted students as frequently.

Universities can utilize the findings of this study to formulate and implement policies aimed at promoting diversity and equality among their graduate students. Furthermore, they can raise awareness among faculty members regarding the benefits, particularly in terms of scientific achievement, that arise from having a diverse research team. Universities can also encourage faculty members to actively consider admitting students from a variety of nationalities.


\begin{backmatter}

\section*{Abbreviations}
  CS, Computer Science; QS, Quacquarelli Symonds; ACM, Association for Computing Machinery; MS, Master of Science; Berkeley, University of California, Berkeley; Caltech, California Institute of Technology; CMU, Carnegie Mellon University; Columbia, Columbia University; Cornell, Cornell University; Georgia Tech, Georgia Institute of Technology; Harvard, Harvard University; McGill, McGill University; MIT, Massachusetts Institute of Technology; NYU, New York University; Princeton, Princeton University; Stanford, Stanford University; U of T, University of Toronto; UBC, University of British Columbia; UChicago, University of Chicago; UCLA, University of California, Los Angeles; UCSD, University of California, San Diego; UIUC, University of Illinois at Urbana-Champaign; UMich, University of Michigan-Ann Arbor; UPenn, University of Pennsylvania; USC, University of Southern California; UT Austin, University of Texas at Austin; UW, University of Washington; Waterloo, University of Waterloo; Yale, Yale University. 

\section*{Availability of data and materials}
  The dataset generated and analyzed during the current study is available in the \textit{Advisor Student Data} repository, \href{https://github.com/kalhorghazal/Advisor-Student-Data}{https://github.com/kalhorghazal/Advisor-Student-Data}.
  
\section*{Competing interests}
  The authors declare that they have no competing interests.

\section*{Funding}
  No funding was received for conducting this study.

\section*{Authors' contributions}
    \textbf{GK:} Data curation, Formal analysis, Methodology, Investigation, Validation, Visualization, Writing- Original draft.
    
\textbf{TZ:} Data curation, Investigation, Validation, Software, Writing- Original draft.

\textbf{BB:} Conceptualization, Project administration, Supervision, Writing- Reviewing and Editing.

\section*{Acknowledgements}
  Thanks to Baharan Khatami for providing some helpful ideas for this research.


\bibliographystyle{bmc-mathphys} 
\bibliography{bmc_article}      


\begin{thebibliography}{39}
\ifx \bisbn   \undefined \def \bisbn  #1{ISBN #1}\fi
\ifx \binits  \undefined \def \binits#1{#1}\fi
\ifx \bauthor  \undefined \def \bauthor#1{#1}\fi
\ifx \batitle  \undefined \def \batitle#1{#1}\fi
\ifx \bjtitle  \undefined \def \bjtitle#1{#1}\fi
\ifx \bvolume  \undefined \def \bvolume#1{\textbf{#1}}\fi
\ifx \byear  \undefined \def \byear#1{#1}\fi
\ifx \bissue  \undefined \def \bissue#1{#1}\fi
\ifx \bfpage  \undefined \def \bfpage#1{#1}\fi
\ifx \blpage  \undefined \def \blpage #1{#1}\fi
\ifx \burl  \undefined \def \burl#1{\textsf{#1}}\fi
\ifx \doiurl  \undefined \def \doiurl#1{\textsf{#1}}\fi
\ifx \betal  \undefined \def \betal{\textit{et al.}}\fi
\ifx \binstitute  \undefined \def \binstitute#1{#1}\fi
\ifx \binstitutionaled  \undefined \def \binstitutionaled#1{#1}\fi
\ifx \bctitle  \undefined \def \bctitle#1{#1}\fi
\ifx \beditor  \undefined \def \beditor#1{#1}\fi
\ifx \bpublisher  \undefined \def \bpublisher#1{#1}\fi
\ifx \bbtitle  \undefined \def \bbtitle#1{#1}\fi
\ifx \bedition  \undefined \def \bedition#1{#1}\fi
\ifx \bseriesno  \undefined \def \bseriesno#1{#1}\fi
\ifx \blocation  \undefined \def \blocation#1{#1}\fi
\ifx \bsertitle  \undefined \def \bsertitle#1{#1}\fi
\ifx \bsnm \undefined \def \bsnm#1{#1}\fi
\ifx \bsuffix \undefined \def \bsuffix#1{#1}\fi
\ifx \bparticle \undefined \def \bparticle#1{#1}\fi
\ifx \barticle \undefined \def \barticle#1{#1}\fi
\ifx \bconfdate \undefined \def \bconfdate #1{#1}\fi
\ifx \botherref \undefined \def \botherref #1{#1}\fi
\ifx \url \undefined \def \url#1{\textsf{#1}}\fi
\ifx \bchapter \undefined \def \bchapter#1{#1}\fi
\ifx \bbook \undefined \def \bbook#1{#1}\fi
\ifx \bcomment \undefined \def \bcomment#1{#1}\fi
\ifx \oauthor \undefined \def \oauthor#1{#1}\fi
\ifx \citeauthoryear \undefined \def \citeauthoryear#1{#1}\fi
\ifx \endbibitem  \undefined \def \endbibitem {}\fi
\ifx \bconflocation  \undefined \def \bconflocation#1{#1}\fi
\ifx \arxivurl  \undefined \def \arxivurl#1{\textsf{#1}}\fi
\csname PreBibitemsHook\endcsname

\bibitem{sharaievska2019we}
\begin{barticle}
\bauthor{\bsnm{Sharaievska}, \binits{I.}},
\bauthor{\bsnm{Kono}, \binits{S.}},
\bauthor{\bsnm{Mirehie}, \binits{M.S.}}:
\batitle{Are we speaking the same language? the experiences of international
  students and scholars in north american higher education}.
\bjtitle{SCHOLE: A Journal of Leisure Studies and Recreation Education}
\bvolume{34}(\bissue{2}),
\bfpage{120}--\blpage{131}
(\byear{2019})
\end{barticle}
\endbibitem

\bibitem{posselt2014toward}
\begin{barticle}
\bauthor{\bsnm{Posselt}, \binits{J.R.}}:
\batitle{Toward inclusive excellence in graduate education: Constructing merit
  and diversity in phd admissions}.
\bjtitle{American Journal of Education}
\bvolume{120}(\bissue{4}),
\bfpage{481}--\blpage{514}
(\byear{2014})
\end{barticle}
\endbibitem

\bibitem{bollinger2007diversity}
\begin{barticle}
\bauthor{\bsnm{Bollinger}, \binits{L.}}:
\batitle{Why diversity matters}.
\bjtitle{The Education Digest}
\bvolume{73}(\bissue{2}),
\bfpage{26}
(\byear{2007})
\end{barticle}
\endbibitem

\bibitem{Smith2020DiversitysPF}
\begin{bchapter}
\bauthor{\bsnm{Smith}, \binits{D.G.}}:
\bctitle{Diversity's promise for higher education: Making it work}.
(\byear{2020})
\end{bchapter}
\endbibitem

\bibitem{maruyama2000university}
\begin{botherref}
\oauthor{\bsnm{Maruyama}, \binits{G.}},
\oauthor{\bsnm{Moreno}, \binits{J.F.}}:
University faculty views about the value of diversity on campus and in the
  classroom.
Does diversity make a difference? Three research studies on diversity in
  college classrooms,
9--35
(2000)
\end{botherref}
\endbibitem

\bibitem{pitman2016understanding}
\begin{barticle}
\bauthor{\bsnm{Pitman}, \binits{T.}}:
\batitle{Understanding ‘fairness’ in student selection: are there
  differences and does it make a difference anyway?}
\bjtitle{Studies in Higher Education}
\bvolume{41}(\bissue{7}),
\bfpage{1203}--\blpage{1216}
(\byear{2016})
\end{barticle}
\endbibitem

\bibitem{Barrera2006MakingUG}
\begin{barticle}
\bauthor{\bsnm{Barrera}, \binits{C.R.}}:
\batitle{Making u.s. graduate education more diverse}.
\bjtitle{Science}
\bvolume{313},
\bfpage{614}--\blpage{614}
(\byear{2006})
\end{barticle}
\endbibitem

\bibitem{bickel1975sex}
\begin{barticle}
\bauthor{\bsnm{Bickel}, \binits{P.J.}},
\bauthor{\bsnm{Hammel}, \binits{E.A.}},
\bauthor{\bsnm{O'Connell}, \binits{J.W.}}:
\batitle{Sex bias in graduate admissions: Data from berkeley: Measuring bias is
  harder than is usually assumed, and the evidence is sometimes contrary to
  expectation.}
\bjtitle{Science}
\bvolume{187}(\bissue{4175}),
\bfpage{398}--\blpage{404}
(\byear{1975})
\end{barticle}
\endbibitem

\bibitem{Maxwell1976}
\begin{barticle}
\bauthor{\bsnm{Maxwell}, \binits{S.E.}},
\bauthor{\bsnm{Jones}, \binits{L.V.}}:
\batitle{Female and male admission to graduate school: An illustrative
  inquiry}.
\bjtitle{Journal of Educational Statistics}
\bvolume{1}(\bissue{1}),
\bfpage{1}--\blpage{37}
(\byear{1976})
\end{barticle}
\endbibitem

\bibitem{attiyeh1997testing}
\begin{botherref}
\oauthor{\bsnm{Attiyeh}, \binits{G.}},
\oauthor{\bsnm{Attiyeh}, \binits{R.}}:
Testing for bias in graduate school admissions.
Journal of Human Resources,
524--548
(1997)
\end{botherref}
\endbibitem

\bibitem{barjak2008international}
\begin{barticle}
\bauthor{\bsnm{Barjak}, \binits{F.}},
\bauthor{\bsnm{Robinson}, \binits{S.}}:
\batitle{International collaboration, mobility and team diversity in the life
  sciences: impact on research performance}.
\bjtitle{Social geography}
\bvolume{3}(\bissue{1}),
\bfpage{23}--\blpage{36}
(\byear{2008})
\end{barticle}
\endbibitem

\bibitem{AlShebli2018}
\begin{botherref}
\oauthor{\bsnm{AlShebli}, \binits{B.K.}},
\oauthor{\bsnm{Rahwan}, \binits{T.}},
\oauthor{\bsnm{Woon}, \binits{W.L.}}:
The preeminence of ethnic diversity in scientific collaboration.
Nature Communications
\textbf{9}(1)
(2018)
\end{botherref}
\endbibitem

\bibitem{llorens2021gender}
\begin{barticle}
\bauthor{\bsnm{Llorens}, \binits{A.}},
\bauthor{\bsnm{Tzovara}, \binits{A.}},
\bauthor{\bsnm{Bellier}, \binits{L.}},
\bauthor{\bsnm{Bhaya-Grossman}, \binits{I.}},
\bauthor{\bsnm{Bidet-Caulet}, \binits{A.}},
\bauthor{\bsnm{Chang}, \binits{W.K.}},
\bauthor{\bsnm{Cross}, \binits{Z.R.}},
\bauthor{\bsnm{Dominguez-Faus}, \binits{R.}},
\bauthor{\bsnm{Flinker}, \binits{A.}},
\bauthor{\bsnm{Fonken}, \binits{Y.}}, \betal:
\batitle{Gender bias in academia: A lifetime problem that needs solutions}.
\bjtitle{Neuron}
\bvolume{109}(\bissue{13}),
\bfpage{2047}--\blpage{2074}
(\byear{2021})
\end{barticle}
\endbibitem

\bibitem{nielsen2018making}
\begin{barticle}
\bauthor{\bsnm{Nielsen}, \binits{M.W.}},
\bauthor{\bsnm{Bloch}, \binits{C.W.}},
\bauthor{\bsnm{Schiebinger}, \binits{L.}}:
\batitle{Making gender diversity work for scientific discovery and innovation}.
\bjtitle{Nature human behaviour}
\bvolume{2}(\bissue{10}),
\bfpage{726}--\blpage{734}
(\byear{2018})
\end{barticle}
\endbibitem

\bibitem{Kamerlin2020}
\begin{botherref}
\oauthor{\bsnm{Kamerlin}, \binits{S.C.L.}}:
When we increase diversity in academia, we all win.
{EMBO} reports
\textbf{21}(12)
(2020)
\end{botherref}
\endbibitem

\bibitem{powell2018these}
\begin{barticle}
\bauthor{\bsnm{Powell}, \binits{K.}}:
\batitle{These labs are remarkably diverse--here's why they're winning at
  science}.
\bjtitle{Nature}
\bvolume{558}(\bissue{7708}),
\bfpage{19}--\blpage{23}
(\byear{2018})
\end{barticle}
\endbibitem

\bibitem{Larsen2005}
\begin{barticle}
\bauthor{\bsnm{Larsen}, \binits{E.A.}},
\bauthor{\bsnm{Stubbs}, \binits{M.L.}}:
\batitle{{INCREASING} {DIVERSITY} {IN} {COMPUTER} {SCIENCE}: {ACKNOWLEDGING},
  {YET} {MOVING} {BEYOND}, {GENDER}}.
\bjtitle{Journal of Women and Minorities in Science and Engineering}
\bvolume{11}(\bissue{2}),
\bfpage{139}--\blpage{170}
(\byear{2005})
\end{barticle}
\endbibitem

\bibitem{Wilson2014}
\begin{barticle}
\bauthor{\bsnm{Wilson}, \binits{C.}}:
\batitle{Hour of code}.
\bjtitle{{ACM} Inroads}
\bvolume{5}(\bissue{4}),
\bfpage{22}--\blpage{22}
(\byear{2014})
\end{barticle}
\endbibitem

\bibitem{Partovi2015}
\begin{barticle}
\bauthor{\bsnm{Partovi}, \binits{H.}}:
\batitle{A comprehensive effort to expand access and diversity in computer
  science}.
\bjtitle{{ACM} Inroads}
\bvolume{6}(\bissue{3}),
\bfpage{67}--\blpage{72}
(\byear{2015})
\end{barticle}
\endbibitem

\bibitem{GarciaHolgado2019}
\begin{bchapter}
\bauthor{\bsnm{Garcia-Holgado}, \binits{A.}},
\bauthor{\bsnm{Vazquez-Ingelmo}, \binits{A.}},
\bauthor{\bsnm{Verdugo-Castro}, \binits{S.}},
\bauthor{\bsnm{Gonzalez}, \binits{C.}},
\bauthor{\bsnm{Gomez}, \binits{M.C.S.}},
\bauthor{\bsnm{Garcia-Penalvo}, \binits{F.J.}}:
\bctitle{Actions to promote diversity in engineering studies: a case study in a
  computer science degree}.
In: \bbtitle{2019 {IEEE} Global Engineering Education Conference ({EDUCON})}.
\bpublisher{{IEEE}}, \blocation{???}
(\byear{2019})
\end{bchapter}
\endbibitem

\bibitem{serrano2009extracting}
\begin{barticle}
\bauthor{\bsnm{Serrano}, \binits{M.{\'A}.}},
\bauthor{\bsnm{Bogun{\'a}}, \binits{M.}},
\bauthor{\bsnm{Vespignani}, \binits{A.}}:
\batitle{Extracting the multiscale backbone of complex weighted networks}.
\bjtitle{Proceedings of the national academy of sciences}
\bvolume{106}(\bissue{16}),
\bfpage{6483}--\blpage{6488}
(\byear{2009})
\end{barticle}
\endbibitem

\bibitem{Blondel_Guillaume_Lambiotte_Lefebvre_2008}
\begin{barticle}
\bauthor{\bsnm{Blondel}, \binits{V.D.}},
\bauthor{\bsnm{Guillaume}, \binits{J.-L.}},
\bauthor{\bsnm{Lambiotte}, \binits{R.}},
\bauthor{\bsnm{Lefebvre}, \binits{E.}}:
\batitle{Fast unfolding of communities in large networks}.
\bjtitle{Journal of Statistical Mechanics: Theory and Experiment}
\bvolume{2008}(\bissue{10}),
\bfpage{10008}
(\byear{2008})
\end{barticle}
\endbibitem

\bibitem{Traag2019}
\begin{botherref}
\oauthor{\bsnm{Traag}, \binits{V.A.}},
\oauthor{\bsnm{Waltman}, \binits{L.}},
\oauthor{\bparticle{van} \bsnm{Eck}, \binits{N.J.}}:
From louvain to leiden: guaranteeing well-connected communities.
Scientific Reports
\textbf{9}(1)
(2019)
\end{botherref}
\endbibitem

\bibitem{diez2012openintro}
\begin{bbook}
\bauthor{\bsnm{Diez}, \binits{D.M.}},
\bauthor{\bsnm{Barr}, \binits{C.D.}},
\bauthor{\bsnm{Cetinkaya-Rundel}, \binits{M.}}:
\bbtitle{OpenIntro Statistics}.
\bpublisher{OpenIntro Boston, MA, USA:}, \blocation{???}
(\byear{2012})
\end{bbook}
\endbibitem

\bibitem{Mann1945}
\begin{barticle}
\bauthor{\bsnm{Mann}, \binits{H.B.}}:
\batitle{Nonparametric tests against trend}.
\bjtitle{Econometrica}
\bvolume{13}(\bissue{3}),
\bfpage{245}
(\byear{1945})
\end{barticle}
\endbibitem

\bibitem{Wei2012}
\begin{bchapter}
\bauthor{\bsnm{Wei}, \binits{D.}},
\bauthor{\bsnm{Li}, \binits{Y.}},
\bauthor{\bsnm{Zhang}, \binits{Y.}},
\bauthor{\bsnm{Deng}, \binits{Y.}}:
\bctitle{Degree centrality based on the weighted network}.
In: \bbtitle{2012 24th Chinese Control and Decision Conference ({CCDC})}.
\bpublisher{{IEEE}}, \blocation{???}
(\byear{2012})
\end{bchapter}
\endbibitem

\bibitem{freeman1978centrality}
\begin{barticle}
\bauthor{\bsnm{Freeman}, \binits{L.C.}}:
\batitle{Centrality in social networks conceptual clarification}.
\bjtitle{Social networks}
\bvolume{1}(\bissue{3}),
\bfpage{215}--\blpage{239}
(\byear{1978})
\end{barticle}
\endbibitem

\bibitem{renyi1959dimension}
\begin{barticle}
\bauthor{\bsnm{R{\'e}nyi}, \binits{A.}}:
\batitle{On the dimension and entropy of probability distributions}.
\bjtitle{Acta Mathematica Academiae Scientiarum Hungarica}
\bvolume{10}(\bissue{1}),
\bfpage{193}--\blpage{215}
(\byear{1959})
\end{barticle}
\endbibitem

\bibitem{topuniversitiesWorldUniversity}
\begin{botherref}
{Q}{S} {W}orld {U}niversity {R}ankings for {C}omputer {S}cience and
  {I}nformation {S}ystems 2021 --- topuniversities.com.
\url{https://www.topuniversities.com/university-rankings/university-subject-rankings/2021/computer-science-information-systems}.
[Accessed 30-Jan-2023]
\end{botherref}
\endbibitem

\bibitem{acmComputingClassification}
\begin{botherref}
{C}omputing {C}lassification {S}ystem --- dl.acm.org.
\url{https://dl.acm.org/ccs}.
[Accessed 30-Jan-2023]
\end{botherref}
\endbibitem

\bibitem{qcriAcuaAudience}
\begin{botherref}
{A}cua: {A}udience, customer, and user analytics --- acua.qcri.org.
\url{https://acua.qcri.org/tool/Name2GAN}.
[Accessed 30-Jan-2023]
\end{botherref}
\endbibitem

\bibitem{li2021gender}
\begin{barticle}
\bauthor{\bsnm{Li}, \binits{B.}},
\bauthor{\bsnm{Jacob-Brassard}, \binits{J.}},
\bauthor{\bsnm{Dossa}, \binits{F.}},
\bauthor{\bsnm{Salata}, \binits{K.}},
\bauthor{\bsnm{Kishibe}, \binits{T.}},
\bauthor{\bsnm{Greco}, \binits{E.}},
\bauthor{\bsnm{Baxter}, \binits{N.N.}},
\bauthor{\bsnm{Al-Omran}, \binits{M.}}:
\batitle{Gender differences in faculty rank among academic physicians: a
  systematic review and meta-analysis}.
\bjtitle{BMJ open}
\bvolume{11}(\bissue{11}),
\bfpage{050322}
(\byear{2021})
\end{barticle}
\endbibitem

\bibitem{liu2018artificial}
\begin{barticle}
\bauthor{\bsnm{Liu}, \binits{J.}},
\bauthor{\bsnm{Kong}, \binits{X.}},
\bauthor{\bsnm{Xia}, \binits{F.}},
\bauthor{\bsnm{Bai}, \binits{X.}},
\bauthor{\bsnm{Wang}, \binits{L.}},
\bauthor{\bsnm{Qing}, \binits{Q.}},
\bauthor{\bsnm{Lee}, \binits{I.}}:
\batitle{Artificial intelligence in the 21st century}.
\bjtitle{IEEE Access}
\bvolume{6},
\bfpage{34403}--\blpage{34421}
(\byear{2018})
\end{barticle}
\endbibitem

\bibitem{schonemann1985artificial}
\begin{barticle}
\bauthor{\bsnm{Sch{\"o}nemann}, \binits{P.H.}}:
\batitle{On artificial intelligence}.
\bjtitle{Behavioral and Brain Sciences}
\bvolume{8}(\bissue{2}),
\bfpage{241}--\blpage{242}
(\byear{1985})
\end{barticle}
\endbibitem

\bibitem{sharma2013h}
\begin{barticle}
\bauthor{\bsnm{Sharma}, \binits{B.}},
\bauthor{\bsnm{Boet}, \binits{S.}},
\bauthor{\bsnm{Grantcharov}, \binits{T.}},
\bauthor{\bsnm{Shin}, \binits{E.}},
\bauthor{\bsnm{Barrowman}, \binits{N.J.}},
\bauthor{\bsnm{Bould}, \binits{M.D.}}:
\batitle{The h-index outperforms other bibliometrics in the assessment of
  research performance in general surgery: a province-wide study}.
\bjtitle{Surgery}
\bvolume{153}(\bissue{4}),
\bfpage{493}--\blpage{501}
(\byear{2013})
\end{barticle}
\endbibitem

\bibitem{cuny2002recruitment}
\begin{barticle}
\bauthor{\bsnm{Cuny}, \binits{J.}},
\bauthor{\bsnm{Aspray}, \binits{W.}}:
\batitle{Recruitment and retention of women graduate students in computer
  science and engineering: results of a workshop organized by the computing
  research association}.
\bjtitle{ACM SIGCSE Bulletin}
\bvolume{34}(\bissue{2}),
\bfpage{168}--\blpage{174}
(\byear{2002})
\end{barticle}
\endbibitem

\bibitem{berg1983men}
\begin{barticle}
\bauthor{\bsnm{Berg}, \binits{H.M.}},
\bauthor{\bsnm{Ferber}, \binits{M.A.}}:
\batitle{Men and women graduate students: Who succeeds and why?}
\bjtitle{The Journal of higher education}
\bvolume{54}(\bissue{6}),
\bfpage{629}--\blpage{648}
(\byear{1983})
\end{barticle}
\endbibitem

\bibitem{okahana2016graduate}
\begin{botherref}
\oauthor{\bsnm{Okahana}, \binits{H.}},
\oauthor{\bsnm{Feaster}, \binits{K.}},
\oauthor{\bsnm{Allum}, \binits{J.}}:
Graduate enrollment and degrees: 2005 to 2015.
Washington, DC: Council of Graduate Schools
(2016)
\end{botherref}
\endbibitem

\bibitem{Sun2019}
\begin{barticle}
\bauthor{\bsnm{Sun}, \binits{Q.}},
\bauthor{\bsnm{Nguyen}, \binits{T.D.}},
\bauthor{\bsnm{Ganesh}, \binits{G.}}:
\batitle{Exploring the study abroad journey: Chinese and indian students in
  u.s. higher education}.
\bjtitle{Journal of International Consumer Marketing}
\bvolume{32}(\bissue{3}),
\bfpage{210}--\blpage{227}
(\byear{2019})
\end{barticle}
\endbibitem

\end{thebibliography}

\newcommand{\BMCxmlcomment}[1]{}

\BMCxmlcomment{

<refgrp>

<bibl id="B1">
  <title><p>Are we speaking the same language? The experiences of international
  students and scholars in North American higher education</p></title>
  <aug>
    <au><snm>Sharaievska</snm><fnm>I</fnm></au>
    <au><snm>Kono</snm><fnm>S</fnm></au>
    <au><snm>Mirehie</snm><fnm>MS</fnm></au>
  </aug>
  <source>SCHOLE: A Journal of Leisure Studies and Recreation
  Education</source>
  <publisher>Taylor \& Francis</publisher>
  <pubdate>2019</pubdate>
  <volume>34</volume>
  <issue>2</issue>
  <fpage>120</fpage>
  <lpage>-131</lpage>
</bibl>

<bibl id="B2">
  <title><p>Toward inclusive excellence in graduate education: Constructing
  merit and diversity in PhD admissions</p></title>
  <aug>
    <au><snm>Posselt</snm><fnm>JR</fnm></au>
  </aug>
  <source>American Journal of Education</source>
  <publisher>University of Chicago Press Chicago, IL</publisher>
  <pubdate>2014</pubdate>
  <volume>120</volume>
  <issue>4</issue>
  <fpage>481</fpage>
  <lpage>-514</lpage>
</bibl>

<bibl id="B3">
  <title><p>Why diversity matters</p></title>
  <aug>
    <au><snm>Bollinger</snm><fnm>L</fnm></au>
  </aug>
  <source>The Education Digest</source>
  <publisher>Prakken Publications, Inc.</publisher>
  <pubdate>2007</pubdate>
  <volume>73</volume>
  <issue>2</issue>
  <fpage>26</fpage>
</bibl>

<bibl id="B4">
  <title><p>Diversity's Promise for Higher Education: Making It
  Work</p></title>
  <aug>
    <au><snm>Smith</snm><fnm>DG</fnm></au>
  </aug>
  <pubdate>2020</pubdate>
</bibl>

<bibl id="B5">
  <title><p>University faculty views about the value of diversity on campus and
  in the classroom</p></title>
  <aug>
    <au><snm>Maruyama</snm><fnm>G</fnm></au>
    <au><snm>Moreno</snm><fnm>JF</fnm></au>
  </aug>
  <source>Does diversity make a difference? Three research studies on diversity
  in college classrooms</source>
  <pubdate>2000</pubdate>
  <fpage>9</fpage>
  <lpage>-35</lpage>
</bibl>

<bibl id="B6">
  <title><p>Understanding ‘fairness’ in student selection: are there
  differences and does it make a difference anyway?</p></title>
  <aug>
    <au><snm>Pitman</snm><fnm>T</fnm></au>
  </aug>
  <source>Studies in Higher Education</source>
  <publisher>Taylor \& Francis</publisher>
  <pubdate>2016</pubdate>
  <volume>41</volume>
  <issue>7</issue>
  <fpage>1203</fpage>
  <lpage>-1216</lpage>
</bibl>

<bibl id="B7">
  <title><p>Making U.S. Graduate Education More Diverse</p></title>
  <aug>
    <au><snm>Barrera</snm><fnm>CR</fnm></au>
  </aug>
  <source>Science</source>
  <pubdate>2006</pubdate>
  <volume>313</volume>
  <fpage>614</fpage>
  <lpage>614</lpage>
</bibl>

<bibl id="B8">
  <title><p>Sex Bias in Graduate Admissions: Data from Berkeley: Measuring bias
  is harder than is usually assumed, and the evidence is sometimes contrary to
  expectation.</p></title>
  <aug>
    <au><snm>Bickel</snm><fnm>PJ</fnm></au>
    <au><snm>Hammel</snm><fnm>EA</fnm></au>
    <au><snm>O'Connell</snm><fnm>JW</fnm></au>
  </aug>
  <source>Science</source>
  <publisher>American Association for the Advancement of Science</publisher>
  <pubdate>1975</pubdate>
  <volume>187</volume>
  <issue>4175</issue>
  <fpage>398</fpage>
  <lpage>-404</lpage>
</bibl>

<bibl id="B9">
  <title><p>Female and Male Admission to Graduate School: An Illustrative
  Inquiry</p></title>
  <aug>
    <au><snm>Maxwell</snm><fnm>SE</fnm></au>
    <au><snm>Jones</snm><fnm>LV</fnm></au>
  </aug>
  <source>Journal of Educational Statistics</source>
  <publisher>American Educational Research Association ({AERA})</publisher>
  <pubdate>1976</pubdate>
  <volume>1</volume>
  <issue>1</issue>
  <fpage>1</fpage>
  <lpage>-37</lpage>
</bibl>

<bibl id="B10">
  <title><p>Testing for bias in graduate school admissions</p></title>
  <aug>
    <au><snm>Attiyeh</snm><fnm>G</fnm></au>
    <au><snm>Attiyeh</snm><fnm>R</fnm></au>
  </aug>
  <source>Journal of Human Resources</source>
  <publisher>JSTOR</publisher>
  <pubdate>1997</pubdate>
  <fpage>524</fpage>
  <lpage>-548</lpage>
</bibl>

<bibl id="B11">
  <title><p>International collaboration, mobility and team diversity in the
  life sciences: impact on research performance</p></title>
  <aug>
    <au><snm>Barjak</snm><fnm>F</fnm></au>
    <au><snm>Robinson</snm><fnm>S</fnm></au>
  </aug>
  <source>Social geography</source>
  <publisher>Copernicus GmbH</publisher>
  <pubdate>2008</pubdate>
  <volume>3</volume>
  <issue>1</issue>
  <fpage>23</fpage>
  <lpage>-36</lpage>
</bibl>

<bibl id="B12">
  <title><p>The preeminence of ethnic diversity in scientific
  collaboration</p></title>
  <aug>
    <au><snm>AlShebli</snm><fnm>BK</fnm></au>
    <au><snm>Rahwan</snm><fnm>T</fnm></au>
    <au><snm>Woon</snm><fnm>WL</fnm></au>
  </aug>
  <source>Nature Communications</source>
  <publisher>Springer Science and Business Media {LLC}</publisher>
  <pubdate>2018</pubdate>
  <volume>9</volume>
  <issue>1</issue>
</bibl>

<bibl id="B13">
  <title><p>Gender bias in academia: A lifetime problem that needs
  solutions</p></title>
  <aug>
    <au><snm>Llorens</snm><fnm>A</fnm></au>
    <au><snm>Tzovara</snm><fnm>A</fnm></au>
    <au><snm>Bellier</snm><fnm>L</fnm></au>
    <au><snm>Bhaya Grossman</snm><fnm>I</fnm></au>
    <au><snm>Bidet Caulet</snm><fnm>A</fnm></au>
    <au><snm>Chang</snm><fnm>WK</fnm></au>
    <au><snm>Cross</snm><fnm>ZR</fnm></au>
    <au><snm>Dominguez Faus</snm><fnm>R</fnm></au>
    <au><snm>Flinker</snm><fnm>A</fnm></au>
    <au><snm>Fonken</snm><fnm>Y</fnm></au>
    <au><cnm>others</cnm></au>
  </aug>
  <source>Neuron</source>
  <publisher>Elsevier</publisher>
  <pubdate>2021</pubdate>
  <volume>109</volume>
  <issue>13</issue>
  <fpage>2047</fpage>
  <lpage>-2074</lpage>
</bibl>

<bibl id="B14">
  <title><p>Making gender diversity work for scientific discovery and
  innovation</p></title>
  <aug>
    <au><snm>Nielsen</snm><fnm>MW</fnm></au>
    <au><snm>Bloch</snm><fnm>CW</fnm></au>
    <au><snm>Schiebinger</snm><fnm>L</fnm></au>
  </aug>
  <source>Nature human behaviour</source>
  <publisher>Nature Publishing Group</publisher>
  <pubdate>2018</pubdate>
  <volume>2</volume>
  <issue>10</issue>
  <fpage>726</fpage>
  <lpage>-734</lpage>
</bibl>

<bibl id="B15">
  <title><p>When we increase diversity in academia, we all win</p></title>
  <aug>
    <au><snm>Kamerlin</snm><fnm>SCL</fnm></au>
  </aug>
  <source>{EMBO} reports</source>
  <publisher>{EMBO}</publisher>
  <pubdate>2020</pubdate>
  <volume>21</volume>
  <issue>12</issue>
</bibl>

<bibl id="B16">
  <title><p>These labs are remarkably diverse--here's why they're winning at
  science</p></title>
  <aug>
    <au><snm>Powell</snm><fnm>K</fnm></au>
  </aug>
  <source>Nature</source>
  <publisher>Nature Publishing Group</publisher>
  <pubdate>2018</pubdate>
  <volume>558</volume>
  <issue>7708</issue>
  <fpage>19</fpage>
  <lpage>-23</lpage>
</bibl>

<bibl id="B17">
  <title><p>{INCREASING} {DIVERSITY} {IN} {COMPUTER} {SCIENCE}:
  {ACKNOWLEDGING}, {YET} {MOVING} {BEYOND}, {GENDER}</p></title>
  <aug>
    <au><snm>Larsen</snm><fnm>EA</fnm></au>
    <au><snm>Stubbs</snm><fnm>ML</fnm></au>
  </aug>
  <source>Journal of Women and Minorities in Science and Engineering</source>
  <publisher>Begell House</publisher>
  <pubdate>2005</pubdate>
  <volume>11</volume>
  <issue>2</issue>
  <fpage>139</fpage>
  <lpage>-170</lpage>
</bibl>

<bibl id="B18">
  <title><p>Hour of code</p></title>
  <aug>
    <au><snm>Wilson</snm><fnm>C</fnm></au>
  </aug>
  <source>{ACM} Inroads</source>
  <publisher>Association for Computing Machinery ({ACM})</publisher>
  <pubdate>2014</pubdate>
  <volume>5</volume>
  <issue>4</issue>
  <fpage>22</fpage>
  <lpage>-22</lpage>
</bibl>

<bibl id="B19">
  <title><p>A comprehensive effort to expand access and diversity in computer
  science</p></title>
  <aug>
    <au><snm>Partovi</snm><fnm>H</fnm></au>
  </aug>
  <source>{ACM} Inroads</source>
  <publisher>Association for Computing Machinery ({ACM})</publisher>
  <pubdate>2015</pubdate>
  <volume>6</volume>
  <issue>3</issue>
  <fpage>67</fpage>
  <lpage>-72</lpage>
</bibl>

<bibl id="B20">
  <title><p>Actions to Promote Diversity in Engineering Studies: a Case Study
  in a Computer Science Degree</p></title>
  <aug>
    <au><snm>Garcia Holgado</snm><fnm>A</fnm></au>
    <au><snm>Vazquez Ingelmo</snm><fnm>A</fnm></au>
    <au><snm>Verdugo Castro</snm><fnm>S</fnm></au>
    <au><snm>Gonzalez</snm><fnm>C</fnm></au>
    <au><snm>Gomez</snm><fnm>MCS</fnm></au>
    <au><snm>Garcia Penalvo</snm><fnm>FJ</fnm></au>
  </aug>
  <source>2019 {IEEE} Global Engineering Education Conference
  ({EDUCON})</source>
  <publisher>{IEEE}</publisher>
  <pubdate>2019</pubdate>
</bibl>

<bibl id="B21">
  <title><p>Extracting the multiscale backbone of complex weighted
  networks</p></title>
  <aug>
    <au><snm>Serrano</snm><fnm>M{\'A}</fnm></au>
    <au><snm>Bogun{\'a}</snm><fnm>M</fnm></au>
    <au><snm>Vespignani</snm><fnm>A</fnm></au>
  </aug>
  <source>Proceedings of the national academy of sciences</source>
  <publisher>National Acad Sciences</publisher>
  <pubdate>2009</pubdate>
  <volume>106</volume>
  <issue>16</issue>
  <fpage>6483</fpage>
  <lpage>-6488</lpage>
</bibl>

<bibl id="B22">
  <title><p>Fast unfolding of communities in large networks</p></title>
  <aug>
    <au><snm>Blondel</snm><fnm>VD</fnm></au>
    <au><snm>Guillaume</snm><fnm>JL</fnm></au>
    <au><snm>Lambiotte</snm><fnm>R</fnm></au>
    <au><snm>Lefebvre</snm><fnm>E</fnm></au>
  </aug>
  <source>Journal of Statistical Mechanics: Theory and Experiment</source>
  <pubdate>2008</pubdate>
  <volume>2008</volume>
  <issue>10</issue>
  <fpage>P10008</fpage>
</bibl>

<bibl id="B23">
  <title><p>From Louvain to Leiden: guaranteeing well-connected
  communities</p></title>
  <aug>
    <au><snm>Traag</snm><fnm>V. A.</fnm></au>
    <au><snm>Waltman</snm><fnm>L.</fnm></au>
    <au><snm>Eck</snm><fnm>N. J.</fnm></au>
  </aug>
  <source>Scientific Reports</source>
  <publisher>Springer Science and Business Media {LLC}</publisher>
  <pubdate>2019</pubdate>
  <volume>9</volume>
  <issue>1</issue>
</bibl>

<bibl id="B24">
  <title><p>OpenIntro statistics</p></title>
  <aug>
    <au><snm>Diez</snm><fnm>DM</fnm></au>
    <au><snm>Barr</snm><fnm>CD</fnm></au>
    <au><snm>Cetinkaya Rundel</snm><fnm>M</fnm></au>
  </aug>
  <publisher>OpenIntro Boston, MA, USA:</publisher>
  <pubdate>2012</pubdate>
</bibl>

<bibl id="B25">
  <title><p>Nonparametric Tests Against Trend</p></title>
  <aug>
    <au><snm>Mann</snm><fnm>HB</fnm></au>
  </aug>
  <source>Econometrica</source>
  <publisher>{JSTOR}</publisher>
  <pubdate>1945</pubdate>
  <volume>13</volume>
  <issue>3</issue>
  <fpage>245</fpage>
</bibl>

<bibl id="B26">
  <title><p>Degree centrality based on the weighted network</p></title>
  <aug>
    <au><snm>Wei</snm><fnm>D</fnm></au>
    <au><snm>Li</snm><fnm>Y</fnm></au>
    <au><snm>Zhang</snm><fnm>Y</fnm></au>
    <au><snm>Deng</snm><fnm>Y</fnm></au>
  </aug>
  <source>2012 24th Chinese Control and Decision Conference ({CCDC})</source>
  <publisher>{IEEE}</publisher>
  <pubdate>2012</pubdate>
</bibl>

<bibl id="B27">
  <title><p>Centrality in social networks conceptual clarification</p></title>
  <aug>
    <au><snm>Freeman</snm><fnm>LC</fnm></au>
  </aug>
  <source>Social networks</source>
  <publisher>North-Holland</publisher>
  <pubdate>1978</pubdate>
  <volume>1</volume>
  <issue>3</issue>
  <fpage>215</fpage>
  <lpage>-239</lpage>
</bibl>

<bibl id="B28">
  <title><p>On the dimension and entropy of probability
  distributions</p></title>
  <aug>
    <au><snm>R{\'e}nyi</snm><fnm>A</fnm></au>
  </aug>
  <source>Acta Mathematica Academiae Scientiarum Hungarica</source>
  <publisher>Springer</publisher>
  <pubdate>1959</pubdate>
  <volume>10</volume>
  <issue>1</issue>
  <fpage>193</fpage>
  <lpage>-215</lpage>
</bibl>

<bibl id="B29">
  <title><p>{Q}{S} {W}orld {U}niversity {R}ankings for {C}omputer {S}cience and
  {I}nformation {S}ystems 2021 --- topuniversities.com</p></title>
  <source>\url{https://www.topuniversities.com/university-rankings/university-subject-rankings/2021/computer-science-information-systems}</source>
  <note>[Accessed 30-Jan-2023]</note>
</bibl>

<bibl id="B30">
  <title><p>{C}omputing {C}lassification {S}ystem --- dl.acm.org</p></title>
  <source>\url{https://dl.acm.org/ccs}</source>
  <note>[Accessed 30-Jan-2023]</note>
</bibl>

<bibl id="B31">
  <title><p>{A}cua: {A}udience, customer, and user analytics ---
  acua.qcri.org</p></title>
  <source>\url{https://acua.qcri.org/tool/Name2GAN}</source>
  <note>[Accessed 30-Jan-2023]</note>
</bibl>

<bibl id="B32">
  <title><p>Gender differences in faculty rank among academic physicians: a
  systematic review and meta-analysis</p></title>
  <aug>
    <au><snm>Li</snm><fnm>B</fnm></au>
    <au><snm>Jacob Brassard</snm><fnm>J</fnm></au>
    <au><snm>Dossa</snm><fnm>F</fnm></au>
    <au><snm>Salata</snm><fnm>K</fnm></au>
    <au><snm>Kishibe</snm><fnm>T</fnm></au>
    <au><snm>Greco</snm><fnm>E</fnm></au>
    <au><snm>Baxter</snm><fnm>NN</fnm></au>
    <au><snm>Al Omran</snm><fnm>M</fnm></au>
  </aug>
  <source>BMJ open</source>
  <publisher>British Medical Journal Publishing Group</publisher>
  <pubdate>2021</pubdate>
  <volume>11</volume>
  <issue>11</issue>
  <fpage>e050322</fpage>
</bibl>

<bibl id="B33">
  <title><p>Artificial intelligence in the 21st century</p></title>
  <aug>
    <au><snm>Liu</snm><fnm>J</fnm></au>
    <au><snm>Kong</snm><fnm>X</fnm></au>
    <au><snm>Xia</snm><fnm>F</fnm></au>
    <au><snm>Bai</snm><fnm>X</fnm></au>
    <au><snm>Wang</snm><fnm>L</fnm></au>
    <au><snm>Qing</snm><fnm>Q</fnm></au>
    <au><snm>Lee</snm><fnm>I</fnm></au>
  </aug>
  <source>IEEE Access</source>
  <publisher>IEEE</publisher>
  <pubdate>2018</pubdate>
  <volume>6</volume>
  <fpage>34403</fpage>
  <lpage>-34421</lpage>
</bibl>

<bibl id="B34">
  <title><p>On artificial intelligence</p></title>
  <aug>
    <au><snm>Sch{\"o}nemann</snm><fnm>PH</fnm></au>
  </aug>
  <source>Behavioral and Brain Sciences</source>
  <publisher>Cambridge University Press</publisher>
  <pubdate>1985</pubdate>
  <volume>8</volume>
  <issue>2</issue>
  <fpage>241</fpage>
  <lpage>-242</lpage>
</bibl>

<bibl id="B35">
  <title><p>The h-index outperforms other bibliometrics in the assessment of
  research performance in general surgery: a province-wide study</p></title>
  <aug>
    <au><snm>Sharma</snm><fnm>B</fnm></au>
    <au><snm>Boet</snm><fnm>S</fnm></au>
    <au><snm>Grantcharov</snm><fnm>T</fnm></au>
    <au><snm>Shin</snm><fnm>E</fnm></au>
    <au><snm>Barrowman</snm><fnm>NJ</fnm></au>
    <au><snm>Bould</snm><fnm>MD</fnm></au>
  </aug>
  <source>Surgery</source>
  <publisher>Elsevier</publisher>
  <pubdate>2013</pubdate>
  <volume>153</volume>
  <issue>4</issue>
  <fpage>493</fpage>
  <lpage>-501</lpage>
</bibl>

<bibl id="B36">
  <title><p>Recruitment and retention of women graduate students in computer
  science and engineering: results of a workshop organized by the computing
  research association</p></title>
  <aug>
    <au><snm>Cuny</snm><fnm>J</fnm></au>
    <au><snm>Aspray</snm><fnm>W</fnm></au>
  </aug>
  <source>ACM SIGCSE Bulletin</source>
  <publisher>ACM New York, NY, USA</publisher>
  <pubdate>2002</pubdate>
  <volume>34</volume>
  <issue>2</issue>
  <fpage>168</fpage>
  <lpage>-174</lpage>
</bibl>

<bibl id="B37">
  <title><p>Men and women graduate students: Who succeeds and why?</p></title>
  <aug>
    <au><snm>Berg</snm><fnm>HM</fnm></au>
    <au><snm>Ferber</snm><fnm>MA</fnm></au>
  </aug>
  <source>The Journal of higher education</source>
  <publisher>Taylor \& Francis</publisher>
  <pubdate>1983</pubdate>
  <volume>54</volume>
  <issue>6</issue>
  <fpage>629</fpage>
  <lpage>-648</lpage>
</bibl>

<bibl id="B38">
  <title><p>Graduate enrollment and degrees: 2005 to 2015</p></title>
  <aug>
    <au><snm>Okahana</snm><fnm>H</fnm></au>
    <au><snm>Feaster</snm><fnm>K</fnm></au>
    <au><snm>Allum</snm><fnm>J</fnm></au>
  </aug>
  <source>Washington, DC: Council of Graduate Schools</source>
  <pubdate>2016</pubdate>
</bibl>

<bibl id="B39">
  <title><p>Exploring the Study Abroad Journey: Chinese and Indian Students in
  U.S. Higher Education</p></title>
  <aug>
    <au><snm>Sun</snm><fnm>Q</fnm></au>
    <au><snm>Nguyen</snm><fnm>TD</fnm></au>
    <au><snm>Ganesh</snm><fnm>G</fnm></au>
  </aug>
  <source>Journal of International Consumer Marketing</source>
  <publisher>Informa {UK} Limited</publisher>
  <pubdate>2019</pubdate>
  <volume>32</volume>
  <issue>3</issue>
  <fpage>210</fpage>
  <lpage>-227</lpage>
</bibl>

</refgrp>
} 





\section*{Figure Legends}

\textbf{Figure \ref{fig:fig1}.} The subgraph is shown before (a) and after (b) the application of the disparity filter algorithm.

\textbf{Figure \ref{fig:fig2}.} Communities are detected using Louvain (c) and Leiden (d) community detection algorithms.

\textbf{Figure \ref{fig:fig3}.} The numbers displayed on the bars indicate the percentage of advisors' gender, while the numbers on the gray section represent the percentage of faculty members in each rank. The width of each bar corresponds to the number of faculty members with a specific rank.

\textbf{Figure \ref{fig:fig5}.} The numbers on each column represent the percentage of different genders in that specific university.

\textbf{Figure \ref{fig:fig9}.} The numbers on the bars represent the percentage of students of each gender, while the numbers on the gray sections indicate the percentage of students in each degree category. The width of each bar is proportional to the number of students in that particular degree category.

\textbf{Figure \ref{fig:fig10}.} The numbers on each column represent the percentage of different genders in that particular university.

\textbf{Figure \ref{fig:fig11}.} The numbers on each bar represent the percentage of different citizenships, and the numbers on the gray section indicate the percentage of students in each degree. The width of each bar is proportional to the number of students in that particular degree.

\end{backmatter}
\end{document}